\documentclass[10pt,onecolumn]{article}
\pdfoutput=1

\usepackage{authblk}

\usepackage[letterpaper, total={6.3in, 9in}]{geometry}
\date{}
\usepackage[utf8]{inputenc}
\usepackage[english]{babel}
\usepackage{csquotes}
\usepackage{lmodern}  
\usepackage{amsmath, amssymb, amsfonts, amsthm}
\usepackage{mathabx}      
\usepackage{stmaryrd}    
\usepackage{cancel}       
\usepackage{physics}
\usepackage{bm}            
\newtheorem{theorem}{Theorem}
\newtheorem{lemma}{Lemma}
\usepackage{graphicx}
\usepackage{tikz}
\usepackage{calc}
\usepackage{caption}
\usepackage{capt-of}
\usepackage{array, booktabs, arydshln, makecell, multicol}
\usepackage{enumitem}
\usepackage{xcolor}
\definecolor{QuantumPurple}{HTML}{53257F}
\usepackage[breaklinks, hidelinks]{hyperref}
\hypersetup{
    colorlinks,
    citecolor=black,
    filecolor=black,
    linkcolor=black,
    urlcolor=black
}
\usepackage{algorithm}
\usepackage{algpseudocode}
\usepackage{algorithmicx}
\usepackage[
    backend=biber, 
    natbib=true,
    style=phys,
    biblabel=brackets,
    giveninits=true,
    abbreviate=false,
    doi=false,
    url=false,
    isbn=false,
    block=space
]{biblatex}
\usepackage[most]{tcolorbox}
\newtcolorbox{egbox}{
    enhanced jigsaw, breakable, oversize,
    colback=QuantumPurple!5!white,
    frame hidden
}
\usepackage[title]{appendix}
\usepackage{marginnote}
\usepackage[normalem]{ulem}
\usepackage{scalerel}
\usepackage{stackengine,wasysym}
\newcommand\reallywidetilde[1]{\ThisStyle{%
  \setbox0=\hbox{$\SavedStyle#1\mathstrut$}%
  \stackengine{.40\LMpt}{$\SavedStyle#1$}{%
    \stretchto{\scaleto{\SavedStyle\mkern.2mu\ACC}{.5150\wd0}}{.3\ht0}%
  }{O}{c}{F}{T}{S}%
}}
\def\ACC{\ooalign{\AC\cr\kern.25\LMpt\AC}}
\newcommand{\textover}[3][l]{%
 \makebox[\widthof{#3}][#1]{#2}%
}
\renewenvironment{abstract}
  {\begin{center}\begin{minipage}{0.85\textwidth}\small}
  {\end{minipage}\end{center}\par\vspace{2ex}}

\usepackage{xcolor}

\author{Miles D.\ Gorman}
\affil{Ming Hsieh Department of Electrical and Computer Engineering, \\
University of Southern California, Los Angeles, California }
 
\begin{document}

\title{Transversal syndrome extraction: A unifying construction for measuring the error syndromes of any stabilizer code}

\maketitle    

\begin{abstract}

  \noindent Fault-tolerant syndrome extraction is fundamental to stabilizer-code-based quantum error correction, yet existing general protocols incur substantial costs in either circuit depth or width. In this work, we design ancilla stabilizer states for a general transversal circuit to construct a family of fault-tolerant syndrome-extraction gadgets that interpolate between Shor and Steane syndrome extraction for any stabilizer code. The result is a configurable trade-off between circuit depth and width. 
\end{abstract}

\section{Introduction}

{Real-world noise can degrade the practical applications of quantum computers. To overcome this challenge, \textit{fault-tolerant} methods using \textit{stabilizer codes} have been proposed \cite{Aharonov1997,Knill1998,ThresholdTheoremConcat,GottesmanConstant,ExpanderCodes,Fawzi2022}. In this paper, we focus on \textit{fault-tolerant syndrome extraction}, a central component of fault tolerant quantum computation, responsible for carefully measuring the effects of noise to inform subsequent correction \cite{FTSErev}.
}

\begin{figure}
    \centering
    \includegraphics[width = \linewidth]{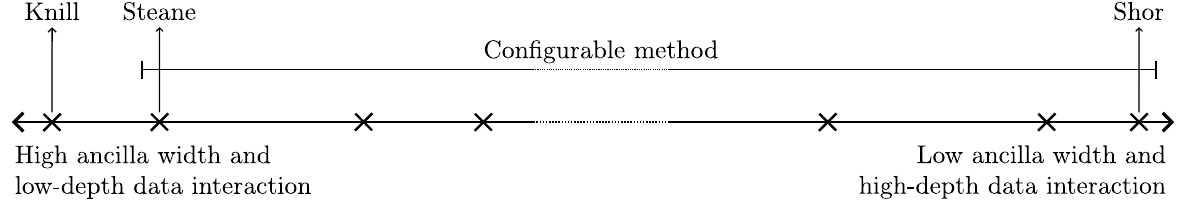}
    \caption{Schematic {comparing resource trade-offs among several syndrome extraction methods.} Each point represents a method’s circuit depth and width for a fixed stabilizer code. Adapted from \cite{gottesman2024surviving}.}
    \label{fig:comparisonofse}
\end{figure}

{
Among the most widely known protocols for fault-tolerant syndrome extraction are those due to {Shor} \cite{Shor96,DiVincenzoShor}, {Steane} \cite{Steane97}, and {Knill} \cite{Knill1,Knill2}. Implementations of these protocols for a given stabilizer code have {fixed, complementary} requirements on circuit depth and width. More recently, configurable constructions have been proposed for a class of stabilizer codes called \textit{Calderbank-Shor-Steane} (CSS) codes \cite{Shinlin1,Shinlin2,ShinlinandYoder}, enabling finer-grained trade-offs between circuit depth and width, as depicted in Figure \ref{fig:comparisonofse}.

In this paper, we present a configurable construction for fault-tolerant syndrome extraction that applies to any stabilizer code, not just CSS codes.} This broadens the resource-optimization options for codes with desirable properties, such as high \textit{encoding rates}. Additionally, in special cases, our scheme recovers Shor-, Steane-, and (in a sense) Knill-style syndrome extraction, unifying them; it offers an alternative route to at least some known configurable syndrome gadgets for CSS codes; and it provides a framework within which a larger family of fault-tolerant syndrome-extraction gadgets may be found.

As a roadmap, this work's background and notation are introduced in Subsections \ref{StabCodes}--\ref{PartiyCheckMatrix}. Subsections \ref{ShorSyndromeExtraction} and \ref{SteaneKnillSyndromeExtraction} use the \textit{stabilizer formalism} to review Shor and Steane/Knill syndrome extraction respectively. Reading the Steane/Knill syndrome extraction subsection is highly recommended, as it presents results that are not widely known but foundational to this work. Subsection \ref{Procedure} outlines our proposed procedure for building syndrome extraction circuits. A detailed examination of these steps follows in Subsections \ref{Explanation}--\ref{subsectionOverhead}. Finally, concluding remarks and directions for future research are provided in Section \ref{Discussion}.

\subsection{Stabilizer codes} \label{StabCodes}

Here, we briefly review stabilizer codes \cite{Roffe,nielsen,gottesmanStabsPaper,gottesman1997stabilizer}.

The \textit{$n$-qubit Pauli group} $\mathcal{G}_n$ is the \textit{group} containing all $n$-fold tensor products of the
identity and \textit{Pauli matrices}, multiplied by phases of $\pm1$ or $\pm i$, where the \textit{Pauli matrices} are

\begin{equation}
    I = \begin{pmatrix} 1 & 0 \\ 0 & 1 \end{pmatrix}, \quad
X = \begin{pmatrix} 0 & 1 \\ 1 & 0 \end{pmatrix}, \quad
Y = \begin{pmatrix} 0 & -i \\ i & 0 \end{pmatrix}, \quad
Z = \begin{pmatrix} 1 & 0 \\ 0 & -1 \end{pmatrix}.
\end{equation}

A \textit{stabilizer group} $\mathcal{S}$ is a \textit{commutative subgroup} of the $n$-qubit Pauli group $\mathcal{G}_n$ that does not contain $-I^{\otimes n}$. It is \textit{closed} under multiplication and can be completely specified by $m\leq n$ independent elements called \textit{generators}, whose products yield every element of the group. We denote a stabilizer group by enclosing its generators in angle brackets, $\langle \dots \rangle$.

Associated with each stabilizer group is a \textit{codespace}, defined as the simultaneous $+1$ eigenspace of its stabilizers. For a stabilizer group with $m$ independent generators acting on $n$ physical qubits, the codespace encodes $k=n-m$ \textit{logical qubits}.

Logical qubits are acted on by \textit{logical operators}. The \textit{logical Pauli operators}, or just \textit{logical operators} for short, are the set $\mathcal{L}$ that contains a choice for the $k$ independent logical Pauli-$X$ operators and $k$ independent logical Pauli-$Z$ operators. The code's logical operators commute with all stabilizers and with each other, except when the $X$ and $Z$ logical operators act on the same logical qubit, in which case they anticommute. Logical $X$ and $Z$ operators acting on logical qubit $i$ are denoted by $\Bar{X_i}$ and $\Bar{Z}_i$ respectively.

The logical Pauli operators of a code determine its error-correction capability. To quantify this, we define the \textit{support} of a Pauli operator as the set of qubits on which it acts nontrivially, and its \textit{weight} as the number of qubits in its support. For a code encoding at least one logical qubit, {the \textit{code distance} $d$ is the minimum weight among all nontrivial logical Pauli operators.} {Such a code,} denoted $\llbracket n,k,d\rrbracket$, corrects every Pauli error of weight at most $t=\lfloor(d-1)/2\rfloor$.

Two important classes of stabilizer codes considered in this paper are stabilizer states and CSS codes. A \textit{stabilizer state} is a stabilizer code with $k=0$ logical qubits, so the stabilizers uniquely specify a single state. CSS codes, on the other hand, are distinguished by the form of their stabilizer generators. Specifically, a stabilizer code is a \textit{CSS code} if its generators can be chosen to consist solely of \textit{$X$-type} and \textit{$Z$-type} operators; an operator is called $X$-type if it is a tensor product of only Pauli-$X$ operators and identities, and $Z$-type operators are defined analogously.

As an example of a stabilizer code, the \textit{five-qubit code} is a non-CSS $\llbracket5,1,3\rrbracket$ code. It has stabilizers
\begin{equation}
    \langle X_1 Z_2 Z_3 X_4, X_2 Z_3 Z_4 X_5, X_1 X_3 Z_4 Z_5, Z_1 X_2 X_4 Z_5 \rangle,
    \label{fivequbitcodestabs}
\end{equation}
and one choice of logical operators is $\Bar{X}_1=X_1X_2X_3X_4X_5$ and $\Bar{Z}_1=Z_1Z_2Z_3Z_4Z_5$. The subscripts denote the single-qubit subsystems on which the supported single-qubit Pauli operators act; identities on the remaining qubits are implicit.

\subsection{Fault-tolerant syndrome extraction} 

The eigenvalues of a code's stabilizer generators on a state can provide a means of correcting errors on that state \cite{Roffe,nielsen,gottesmanStabsPaper,gottesman1997stabilizer}. The process of measuring these eigenvalues is called \textit{syndrome extraction}, and the resulting vector of measurement outcomes is called the \textit{syndrome}. Since stabilizers are Pauli operators, their eigenvalues are either $+1$ or $-1$. By convention, these are mapped to $0$ and $1$ in the syndrome, respectively,  meaning the syndrome is an element of $\mathbb{Z}_2^m$.

Because quantum operations are noisy, errors can occur during a circuit; these are called \textit{faults}. Faults can propagate through the circuit and cause errors on additional qubits, potentially undermining the benefits of error correction. To limit their effects, circuits must be \textit{fault tolerant}.

For a code capable of correcting up to $t$ errors, the following conditions are sufficient for a syndrome-extraction circuit, used within an error correction protocol, to be fault tolerant:

\begin{enumerate}
    \item If an encoded input state has $u$ errors, and $v$ faults occur, with $u+v\leq t$, fault-free error correction of the output state must recover the same encoded state as fault-free error correction of the input.
    
    \item For an encoded input state with an arbitrary number of errors, if at most $v\leq t$ faults occur, the output after error correction must differ from a state in the codespace by an error of weight at most $v$.
\end{enumerate}

{
\noindent The first condition ensures that logical errors are suppressed to higher orders in the physical error rate. {Together with the second condition,} this helps establish fault-tolerant quantum computation, in which logical error rates can be suppressed arbitrarily at a modest asymptotic cost, provided the physical error rate is below a scheme-dependent \textit{threshold} value \cite{ThresholdTheoremConcat, gottesman2009introductionquantumerrorcorrection}.}

A common building block of fault-tolerant circuits, particularly relevant to this paper, is the \textit{transversal two-qubit gate}, which acts on pairs of qubits belonging to separate encoded subsystems. This structure prevents errors from propagating between qubits within the same subsystem, thereby limiting their spread and facilitating fault tolerance \cite{gottesman2024surviving, gottesman2009introductionquantumerrorcorrection}.

\subsection{Pauli propagation} \label{PauliPropagation}

{\textit{Pauli propagation} provides an efficient framework for computing quantum state dynamics in certain circuits \cite{nielsen,gottesman1997stabilizer}. It suffices for all analyses in this paper, and we introduce it below.}

Consider a stabilizer or logical operator $P\in \mathcal{S} \cup \mathcal{L} \subset \mathcal{G}_n$ of an $n$-qubit encoded state
$\ket{\psi}_L$. If the state $\ket{\psi}_L$ is acted on by the operator $P$, the resulting state is $P\ket{\psi}_L$.
Applying a unitary $U$ to that state $P\ket{\psi}_L$ gives

\begin{equation}
UP\ket{\psi}_L=UPU^\dagger U\ket{\psi}_L.
\end{equation}

\noindent Therefore, since $P$ is a logical operator or stabilizer of state $\ket{\psi}_L$, $UPU^\dagger$ is a logical operator or stabilizer of the {transformed} state $U\ket{\psi}_L$; colloquially, we say $P$ \textit{propagates} to $UPU^\dagger$. {This allows us to describe how the codespace and its logical operators change under a unitary circuit. What is more, because this mapping} is a \textit{group homomorphism}, {it suffices to track the propagation of a generating set rather than every group element \cite{nielsen,gottesman1997stabilizer}.}

We will only be concerned with how Pauli operators {propagate} through \textit{two-qubit controlled Pauli gates}, defined as
\begin{equation}
    \text{C}[P]_{i,j} \equiv (\ket{0}\bra{0})_i \otimes I_j + (\ket{1}\bra{1})_i \otimes P_j,
\end{equation}
where $P\in\{X,Y,Z\}$ and there are implicit identities on all other qubits in the system. The relevant circuit identities are
\begin{align}
    \text{C}[P]_{i,j} X_k \text{C}[P]_{i,j}^\dagger =& \begin{cases}   X_i P_j   & \text{for } k=i\\
    Z_i X_j   & \text{for } k=j \text{ and } P\in\{Y,Z\} \\
                                     X_k & \text{otherwise}                         \end{cases},\\
    \text{C}[P]_{i,j} Z_k \text{C}[P]_{i,j}^\dagger =& \begin{cases}   Z_i Z_j   & \text{for } k=j \text{ and } P\in\{X,Y\}\\
                           Z_k & \text{otherwise}
                         \end{cases},
\end{align}
which encapsulates how $Y_k=iX_kZ_k$, and hence any Pauli operator, propagates through these gates.

\subsection{Matrix representation} \label{PartiyCheckMatrix}

In this subsection, we introduce a convenient matrix representation for a system's stabilizers and logical operators.

Any stabilizer or logical operator may be written, up to a known phase, as
\begin{equation}
    (-1)^s X^{x_1}_1 Z^{z_1}_1 X^{x_2}_2 Z^{z_2}_2 \cdots X^{x_n}_n Z^{z_n}_n \cong     \left( \begin{array}{@{}c@{}|c@{}|c@{}}
    \begin{matrix}
        s
    \end{matrix}
        &
       \begin{matrix}
          x_1 & x_2 & \dots & x_n
       \end{matrix}
       &
       \begin{matrix}
          z_1 & z_2 & \dots & z_n
       \end{matrix}
       \end{array} \right) \in \mathbb{Z}_2^{2n+1},
    \label{ParityCheckRow}
\end{equation}
where ${x},{z} \in \mathbb{Z}_2^n$ and $s\in \mathbb{Z}_2$. For convenience, and without loss of generality, we omit known phases throughout, retaining only the bit $s$ to account for unknown $\pm 1$ phases on stabilizers arising from errors. {Disregarding known phases also simplifies operator multiplication in this representation to addition of the corresponding vectors modulo $2$, denoted by $\oplus$.}

An encoded state with $n$ physical qubits and $m$ generators can be partially represented by a matrix $\mathcal{S}\in \mathbb{Z}_2^{m\times (2n+1)}$, where each row is the vector representation of a stabilizer generator. Similarly, the corresponding code's $2k$ logical operators can be represented by a matrix $\mathcal{L}\in \mathbb{Z}_2^{2k\times 2n}$. Note that we will often abuse notation and refer to Pauli operators, their binary vector representations, and their corresponding matrix rows interchangeably.

It is useful to further decompose an encoded {state's} stabilizers and logical operators into their Pauli-$X$ and Pauli-$Z$ components, {corresponding to} the left- and right-hand sides of their vector representations. {More precisely, the Pauli-$X$ and Pauli-$Z$ components of the code's stabilizers and logical operators are denoted by} $\mathcal{S}_X,\mathcal{S}_Z\in \mathbb{Z}_2^{m\times n}$ and $\mathcal{L}_X,\mathcal{L}_Z\in \mathbb{Z}_2^{2k\times n}$
respectively and defined by the relations

\begin{align}
\label{stabilzierdecomp}
    \mathcal{S} =
        \left( \begin{array}{@{}c|c@{}|c@{}}
        s&
       \begin{matrix}
          \mathcal{S}_X
       \end{matrix} &
       \begin{matrix}
          \mathcal{S}_Z
       \end{matrix}
       \end{array} \right),\\
        \mathcal{L} =
        \left( \begin{array}{@{}c@{}|c@{}}
       \begin{matrix}
          \mathcal{L}_X
       \end{matrix}
       &
       \begin{matrix}
          \mathcal{L}_Z
       \end{matrix}
       \end{array} \right),
\end{align}

\noindent where $s\in \mathbb{Z}_2^m$ is a vector {effectively describing the encoded state's syndrome}. {When the entries of column $s$ are fixed, we may omit it, leaving the code's \textit{parity-check matrix} $(\mathcal{S}_X\mid\mathcal{S}_Z)$.}

We combine these representations to express the stabilizers and logical operators of an encoded state collectively as

    \begin{equation}
       \left( 
       \begin{array}{@{}c|c@{}|c@{}}
            s&
           \begin{matrix}
               \mathcal{S}_X
           \end{matrix}
           
           &
           
           \begin{matrix}
              \mathcal{S}_Z
           \end{matrix}

          \\
          \hline
          &
            \begin{matrix}
                   \textover[c]{$\mathcal{L}_X$}{$\mathcal{S}_X$}
               \end{matrix}
               
               &
               
               \begin{matrix}
                   \textover[c]{$\mathcal{L}_Z$}{$\mathcal{S}_Z$}
               \end{matrix}
       
     \end{array} 
     \right)
   \end{equation}

\noindent where the stabilizer generators and logical operators of the code are separated by a horizontal line. 
The absence of such a horizontal line indicates that the code has no nontrivial logical operators.

{
To illustrate this representation, recall the five-qubit code from Eq.\ \ref{fivequbitcodestabs}.} It has a stabilizer

\begin{equation}
    S =X_1 Z_2 Z_3 X_4 \cong \left( \begin{array}{@{}c@{}|c@{}|c@{}}
    \begin{matrix}
          0
       \end{matrix}
       &
       \begin{matrix}
          1 & 0 & 0 & 1 & 0
       \end{matrix}
       &
       \begin{matrix}
          0 & 1 & 1 &0 & 0
       \end{matrix}
       \end{array} \right),
\end{equation}

\noindent which can be decomposed into $S_X = \left(
       \begin{matrix}
          1 & 0 & 0 & 1 & 0
       \end{matrix}
    \right)$ and $   S_Z=     \left(
       \begin{matrix}
          0 & 1 & 1 &0 & 0
       \end{matrix}
    \right)$.
More completely, the five-qubit code {can be} described by the matrix 

   \begin{equation}
       \left( \begin{array}{@{}c@{}|c@{}|c@{}}
       \begin{matrix}
           0\\0\\0\\0
       \end{matrix}
       &
       \begin{matrix}
           1 & 0 & 0 & 1 & 0\\
           0 & 1 & 0 & 0 & 1\\
           1 & 0 & 1 & 0 & 0\\
           0 & 1 & 0 & 1 & 0
       \end{matrix}
       &
       \begin{matrix}
          0 & 1 & 1 &0 & 0\\
           0 & 0 & 1 &1 & 0\\
           0 & 0 & 0 &1 & 1\\
           1 & 0 & 0 &0 & 1
       \end{matrix}
          \\\hline
         
        \begin{matrix}
            \phantom{0}\\\phantom{0}
        \end{matrix}
        &
           \begin{matrix}
           1 & 1 & 1 & 1 & 1\\
           0 & 0 & 0 & 0 & 0
       \end{matrix}
       &
       \begin{matrix}
           0 & 0 & 0 & 0 & 0\\
           1 & 1 & 1 & 1 & 1
       \end{matrix}
     \end{array} \right).
   \end{equation}

\subsection{Shor syndrome extraction} \label{ShorSyndromeExtraction}

We now review Shor syndrome extraction, a method for fault-tolerant syndrome extraction \cite{Shor96,DiVincenzoShor,FTSErev}.

{
The procedure for measuring a weight-$w$ stabilizer $S\in\mathcal{S}$ is as follows:
}

\begin{figure}
 \centering
 \includegraphics{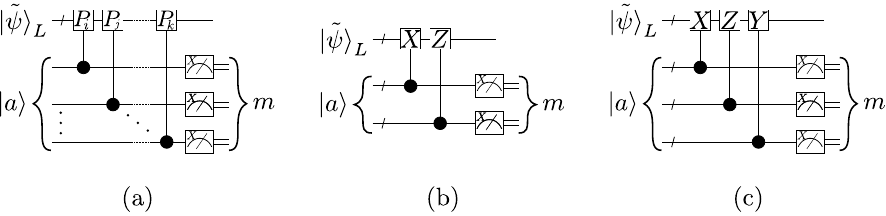}
 \caption{Circuit diagrams for (a) Shor syndrome extraction; (b) Steane syndrome extraction, Knill syndrome extraction, and our proposed procedure{, up to circuit identities, the deletion of certain ancilla qubits, and the choice of ancilla state $\ket{a}$; and (c) a more general version of our procedure, again up to deleting certain ancilla qubits. In all cases, $\ket{\Tilde{\psi}}_L$ represents an $n$-qubit input state initially prepared in the codespace and possibly corrupted by subsequent errors, $\ket{a}$ denotes a scheme-dependent ancilla state,} and ${m}$ contains the {single-qubit $X$-basis} measurement outcomes. {A slash on a wire denotes a \textit{block} of $n$ qubits, and each controlled-Pauli gate in (b) and (c) acts transversally between corresponding qubits in the data and ancilla blocks. In (a), the controlled-Pauli gates correspond to the support of the measured stabilizer $P_i P_j \dots P_k$, where each $P_l\in\mathcal{G}_n$ is a weight-1 Pauli operator acting nontrivially on qubit $l$.}}
 \label{fig:ShorAndSteane}
\end{figure}

\begin{enumerate}
    \item {Prepare a $w$-qubit \textit{cat state}, defined by its stabilizer group  \begin{equation}
        \langle X_1 X_2 \dots X_w, Z_1 Z_2, Z_2 Z_3,\dots, Z_{w-1} Z_w \rangle.
    \end{equation}
    For $w=1$, the $Z$-type generators are absent.}

    \item {Apply the gates in Figure \ref{fig:ShorAndSteane}a, where the stabilizer being measured is decomposed into a product of single-qubit Pauli operators, i.e., $S=P_iP_j\dots P_k$ where each $P_l\in\mathcal{G}_n$ is a weight-1 Pauli operator acting nontrivially on qubit $l$.}

    \item Calculate the stabilizer's syndrome bit by {adding the measurement outcomes modulo 2.}

    \item {Repeat for different stabilizers $S\in\mathcal{S}$ until at least a generating set has been measured.}
\end{enumerate}

{
To understand why this procedure works, consider} how the $X_1 X_2 \dots X_w$ ancilla stabilizer propagates through the circuit, recalling that the subscripts $1$, $2$, $\dots$, $w$ label ancilla qubits. Namely, view the data as being stabilized by $(-1)^sS_{\underline{{d}}}$ where $s\in\mathbb{Z}_2$ is that stabilizer's syndrome bit we are trying to measure and $\underline{{d}}$ specifies that this operator acts nontrivially only on the data qubits. The ancilla stabilizer $X_1 X_2 \dots X_w$, up to measurement, propagates to $S_{\underline{{d}}}X_1 X_2 \dots X_w$, while the data stabilizer $(-1)^sS_{\underline{{d}}}$ remains unchanged. Using that the stabilizer group of the system is closed, we can thus see that 

\begin{equation}
(-1)^sS^2_{\underline{{d}}}X_1 X_2 \dots X_w=(-1)^sX_1 X_2 \dots X_w
\end{equation}

\noindent is also a stabilizer of the composite system. Then, by measuring the ancilla in the $X$ basis and retrieving outcomes $m_1$, $m_2$, $\dots$, $m_w$, the system's state collapses such that it is stabilized by $(-1)^{m_1}X_1$, $(-1)^{m_2}X_2$, $\dots$, $(-1)^{m_w}X_w$, and $(-1)^sX_1 X_2 \dots X_w$, with which the measurements commute. Thus

\begin{equation}
    (-1)^s(-1)^{m_1}X^2_1 (-1)^{m_2}X^2_2 \dots (-1)^{m_w}X^2_w=(-1)^{s \oplus m_1\oplus m_2 \oplus \dots \oplus m_w}I
    \label{MEASURMENTSEQ}
\end{equation}

\noindent is also a stabilizer of the system, and because $-I$ cannot be a stabilizer, we must have that the syndrome bit is $s=m_1\oplus m_2 \oplus \dots \oplus m_w$.

As for the $\{Z_1 Z_2, Z_2 Z_3,\dots, Z_{w-1} Z_w\}$ stabilizer generators, the key observation is that they remain $Z$-type throughout the circuit. Consequently, the $X$-basis measurements do not reveal any information on the data state beyond the intended syndrome bit.

\subsection{Steane/Knill syndrome extraction} \label{SteaneKnillSyndromeExtraction}

In this subsection, we review a generalized version of Steane syndrome extraction \cite{GeneralSteane}, distinguished from standard Steane syndrome extraction \cite{Steane97} because it works for all stabilizer codes. Generalized Steane syndrome extraction is equivalent to Knill syndrome extraction \cite{Knill1,Knill2} up to circuit identities \cite{GeneralSteane}, and it reduces to standard Steane syndrome extraction when used with CSS codes. We will henceforth refer to this generalized Steane syndrome extraction simply as \textit{Steane syndrome extraction}.

{
For a stabilizer code with $m\times2n$ parity-check matrix
$\mathcal{S}\equiv(\mathcal{S}_X|\mathcal{S}_Z)$ and logical operators
$\mathcal{L}\equiv(\mathcal{L}_X|\mathcal{L}_Z)$, the procedure is as follows:
}

\begin{enumerate}
    \item Initialize a $2n$-qubit ancilla state stabilized by

    \begin{equation}
       \left( \begin{array}{@{}c@{}|c@{}}
       \begin{matrix}
          \mathcal{S}_X & \mathcal{S}_Z \\
          0 & 0 \\
          0 & 0
       \end{matrix}
       &
       \begin{matrix}
          0 & \mathcal{S}_X \\
          \mathcal{S}_Z & \mathcal{S}_X \\
          \mathcal{L}_Z & \mathcal{L}_X
       \end{matrix}
       \end{array} \right).
        \label{SteaneAncillaState}
   \end{equation}

   \item Run the circuit shown in Figure \ref{fig:ShorAndSteane}b.

   \item If the measurement outcomes are ${m}\in\mathbb{Z}_2^{2n}$, the syndrome is $s=\mathcal{S}{m}$ modulo 2, up to addition modulo 2 of a known vector accounting for the known phases suppressed in the matrix.
\end{enumerate}

To see why this procedure works, consider the stabilizers and logical operators of the initial data-ancilla state,

    \begin{equation}
       \left( \begin{array}{@{}c@{}|c@{}|c@{}}
       \begin{matrix}
        s\\0\\0\\0   
       \end{matrix}
       &
       \begin{matrix}
           \mathcal{S}_X & 0 & 0\\
          0 & \mathcal{S}_X & \mathcal{S}_Z \\
          0 & 0 & 0 \\
          0 & 0 & 0 \\
       \end{matrix}
       &
       \begin{matrix}
          \mathcal{S}_Z & 0 & 0\\
          0 & 0 & \mathcal{S}_X \\
          0 & \mathcal{S}_Z & \mathcal{S}_X \\
          0 & \mathcal{L}_Z & \mathcal{L}_X \\
       \end{matrix}
          \\ \hline
          \begin{matrix}
              \phantom{0}
          \end{matrix}
          &
                       \begin{matrix}
           \textover[c]{$\mathcal{L}_X$}{$\mathcal{S}_X$} & \textover[c]{$0$}{$\mathcal{S}_X$} & \textover[c]{$0$}{$\mathcal{S}_Z$}\\
       \end{matrix}
       &
       \begin{matrix}
           \textover[c]{$\mathcal{L}_Z$}{$\mathcal{S}_X$} & \textover[c]{$0$}{$\mathcal{S}_X$} & \textover[c]{$0$}{$\mathcal{S}_Z$}\\
       \end{matrix}
     \end{array} \right),
   \end{equation}

\noindent {where the second and fifth columns denote the data qubits and the remaining columns (besides the first column, which holds the syndrome $s\in \mathbb{Z}^m_2$ we aim to measure) denote the ancilla qubits. Up to measurements, these operators propagate through the circuit to}

    \begin{equation}
       \left( \begin{array}{@{}c@{}|c@{}|c@{}}
       \begin{matrix}
        s\\0\\0\\0   
       \end{matrix}
       &
       \begin{matrix}
           \mathcal{S}_X & 0 & 0\\
          \mathcal{S}_X & \mathcal{S}_X & \mathcal{S}_Z \\
          0 & 0 & 0 \\
          0 & 0 & 0
       \end{matrix}
       &
       \begin{matrix}
          \mathcal{S}_Z & \mathcal{S}_Z & \mathcal{S}_X\\
          \mathcal{S}_Z & 0 & 0 \\
          0 & \mathcal{S}_Z & \mathcal{S}_X \\
          0 & \mathcal{L}_Z & \mathcal{L}_X
       \end{matrix}\\
                  \hline
                  \begin{matrix}
        \phantom{0} 
       \end{matrix}
       &
                  \begin{matrix}
          \textover[c]{$\mathcal{L}_X$}{$\mathcal{S}_X$} & \textover[c]{$0$}{$\mathcal{S}_X$} & \textover[c]{$0$}{$\mathcal{S}_Z$}\\
       \end{matrix}
       &
       \begin{matrix}
          \mathcal{L}_Z & \mathcal{L}_Z & \mathcal{L}_X\\
       \end{matrix}
       \end{array} \right)
\cong
       \left( \begin{array}{@{}c@{}|c@{}|c@{}}
       \begin{matrix}
        s\\s\\0\\0   
       \end{matrix}
       &
       \begin{matrix}
            \mathcal{S}_X & 0 & 0\\
          0 & \mathcal{S}_X & \mathcal{S}_Z \\
          0 & 0 & 0 \\
          0 & 0 & 0
       \end{matrix}
       &
       \begin{matrix}
          \mathcal{S}_Z & 0 & 0\\
          0 & 0 & 0 \\
          0 & \mathcal{S}_Z & \mathcal{S}_X \\
          0 & \mathcal{L}_Z & \mathcal{L}_X
       \end{matrix}\\
                  \hline
                  \begin{matrix}
        \phantom{0} 
       \end{matrix}
       &
                       \begin{matrix}
           \textover[c]{$\mathcal{L}_X$}{$\mathcal{S}_X$} & \textover[c]{$0$}{$\mathcal{S}_X$} & \textover[c]{$0$}{$\mathcal{S}_Z$}\\
       \end{matrix}
       &
       \begin{matrix}
           \textover[c]{$\mathcal{L}_Z$}{$\mathcal{S}_X$} & \textover[c]{$0$}{$\mathcal{S}_X$} & \textover[c]{$0$}{$\mathcal{S}_Z$}\\
       \end{matrix}
     \end{array} \right),
   \end{equation}

\noindent {where the {equivalence follows from multiplying generators.}} 

{Consider the role played by the Pauli-$X$ support of the ancilla's stabilizers. As in Shor syndrome extraction, these Pauli-$X$ terms propagate through the circuit such that they acquire {the stabilizers whose syndrome bits are to be measured}, effectively transferring the syndrome bits onto said ancilla stabilizers. Moreover, the $Z$ support of the ancilla stabilizers is chosen to cancel any $Z$ terms propagating from the data, leaving each ancilla stabilizer either $X$-type or $Z$-type after the gates. Measuring the ancilla in the $X$ basis therefore reveals only the syndrome; by the same reasoning as in Shor's scheme (Eq.\ \ref{MEASURMENTSEQ}), the syndrome is obtained by adding certain measurement outcomes modulo 2. This cancellation, enabled by the $Z$ support of the ancilla stabilizers, also leaves the data stabilizers and logical operators unchanged, thereby leaving the data state $\Tilde{\ket{\psi}}$ unchanged by the circuit \cite{FTSErev}.
}

\section{Results} \label{Results}

In this section, we present our method for syndrome extraction. The first subsection outlines the steps of the syndrome extraction gadget; the second explains why it works; {the third relates our scheme to Shor and Steane syndrome extraction and to an existing configurable method for a CSS code}; the fourth explains its fault tolerance; and the fifth analyzes its overhead and performance.

\subsection{Procedure} \label{Procedure}

The basic idea is to measure sets of stabilizers together, using carefully constructed ancilla states and the circuit shown in Figure \ref{fig:ShorAndSteane}b. Specifically, for a stabilizer code with $m\times2n$ parity-check matrix
$\mathcal{S}\equiv\left( \begin{array}{@{}c|c@{}} \mathcal{S}_X &  \mathcal{S}_Z  \end{array} \right)$
and logical operators
$\mathcal{L}\equiv\left( \begin{array}{@{}c|c@{}} \mathcal{L}_X &  \mathcal{L}_Z  \end{array} \right)$:

\begin{enumerate}

\item Partition the stabilizer generators $\mathcal{S}=\mathcal{S}^{(M)}\cup \mathcal{S}^{(L)}$ where $\mathcal{S}^{(M)}\subseteq \mathcal{S}$ contains the stabilizers we wish to measure the syndrome bit of.

\item  Choose $\mathcal{S}^{(LM)} \cup \mathcal{S}^{(LL)}$ as a new generating set for $\langle \mathcal{S}^{(L)} \rangle$, where:

    \begin{enumerate}
        \item Every zero column of $\mathcal{S}^{(M)}$ is also a zero column of $\mathcal{S}^{(LM)}$;
        \item Every row of ${\mathcal{S}}^{(LL)}$ has an entry of 1 in a column that no other row of $\mathcal{S}^{(M)}\cup{\mathcal{S}}^{(LL)}$ has a 1 in.
    \end{enumerate}
    Appendix \ref{stabilizerform} provides an algorithm to achieve this for any stabilizer code.

    \item Consider the ancilla of generalized Steane syndrome extraction, which has parity-check matrix

    \begin{equation}
       \left( \begin{array}{@{}c@{}|c@{}}
       \begin{matrix}
          \mathcal{S}_X & \mathcal{S}_Z \\
          0 & 0 \\
          0 & 0 \\

       \end{matrix}
       &
       \begin{matrix}
          0 & \mathcal{S}_X \\
          \mathcal{S}_Z & \mathcal{S}_X \\
          \mathcal{L}_Z & \mathcal{L}_X\\
       \end{matrix}
       \end{array} \right)
=     
       \left( \begin{array}{@{}c@{}|c@{}}
       \begin{matrix}
          {\mathcal{S}}_X^{(LL)} & {\mathcal{S}}_Z^{(LL)} \\
          {\mathcal{S}}_X^{(LM)} & {\mathcal{S}}_Z^{(LM)} \\
          \mathcal{S}_X^{(M)} & \mathcal{S}_Z^{(M)} \\
          0 & 0 \\
          0 & 0 \\
       \end{matrix}
       &
       \begin{matrix}
          0 & {\mathcal{S}}_X^{(LL)} \\
          0 & {\mathcal{S}}_X^{(LM)} \\
          0 & \mathcal{S}_X^{(M)} \\
          \mathcal{S}_Z & \mathcal{S}_X \\
          \mathcal{L}_Z & \mathcal{L}_X
       \end{matrix}
       \end{array} \right).
        \label{AncillaStabsProcedure}
   \end{equation}

     Remove the rows
    $( \begin{array}{@{}c  c@{}} \begin{matrix} {\mathcal{S}}_X^{(LL)} &  {\mathcal{S}}_Z^{(LL)} \end{matrix} \mid& \begin{matrix} 0 & {\mathcal{S}}_X^{(LL)}  \end{matrix} \end{array} \!)$ and identify the zero columns in $\mathcal{S}^{(M)}$.
    If column $j\in\{1,2,\dots,2n\}$ is a zero column of $\mathcal{S}^{(M)}$, {set column $2n+j$ of the parity-check matrix above to zero, and include the generator $Z_j$ by appending a row whose only nonzero entry is a 1 in column $2n+j$.}
Repeat for all such zero columns and prepare an ancilla state with those stabilizers.

    \item Apply the gates in Figure \ref{fig:ShorAndSteane}b with the above ancilla, but deleting from the ancilla any qubits (and gates acting on those qubits) that are stabilized by single-qubit Pauli-$Z$s.

    \item  If the measurement outcomes are ${m}\in\mathbb{Z}_2^{r}$ for some $r\in\mathbb{N}$, the syndrome, up to addition modulo 2 of a known vector accounting for the phases suppressed in our matrix calculations, is $s=(\reallywidetilde{\mathcal{S}^{(M)} \cup {\mathcal{S}}^{(LM)}}){m}$ where $(\reallywidetilde{\mathcal{S}^{(M)} \cup {\mathcal{S}}^{(LM)}})$ is the parity-check matrix of $\mathcal{S}^{(M)} \cup {\mathcal{S}}^{(LM)}$ but with the columns corresponding to deleted qubits removed.

    \item Repeat for different choices of $\mathcal{S}^{(M)}$ until there is at least one syndrome bit for every stabilizer generator. 

\end{enumerate}

\noindent A pedagogical example of this procedure applied to the five-qubit code is provided in Appendix \ref{secexample}.

Note that the procedure can be modified to make use of controlled-$Y$ gates. In particular, if {a subset of} qubits in $\mathcal{S}^{(M)}$ and ${\mathcal{S}}^{(LM)}$ have only Pauli-$Y$s or identities acting on them, then for each of these qubits, we can merge the $\text{C}[X]$ and $\text{C}[Z]$ gates targeting them into a $\text{C}[Y]$ gate controlled on one of their ancilla qubits, and remove the other ancilla qubit.

\subsection{Validity of our procedure} \label{Explanation}

The procedure involves adding and removing stabilizers from Steane syndrome extraction's ancilla, as well as deleting certain qubits and gates. Generally, such modifications do not preserve a valid stabilizer group, let alone one that uniquely specifies an ancilla state enabling extraction of only the intended syndrome bits. In this subsection, we outline the results establishing that our procedure satisfies these requirements for any stabilizer code; the corresponding proofs are deferred to the appendix because of their length.

Steps 1 and 2 of the procedure put the stabilizers in a specific form. This form can be achieved for any stabilizer code by linearly combining rows of $\mathcal{S}^{(L)}$ according to the algorithm described in Appendix \ref{stabilizerform}.

Steps 3 and 4 alter the circuit and ancilla state. By utilizing commutation relations and the structure of the stabilizers introduced in Step 2, Appendix \ref{validstabs} shows that the stabilizer group of the ancilla state is valid. Appendix \ref{ProofSyndromeExtraction} shows that the syndrome bits of the stabilizers $\mathcal{S}^{(M)} \cup {\mathcal{S}}^{(LM)}$ are indeed extracted and given according to Step 5. This is done by considering how the data and ancilla's stabilizers propagate through the circuit.

{
The results above also hold after implementing the modification regarding controlled-Y gates, as shown in Appendix \ref{CYgateresults}, using essentially the same arguments.
}

\subsection{ Connections to existing syndrome-extraction schemes}\label{subsecshorsteaneknillconnection}

{
In this subsection, we identify the conditions under which our scheme reproduces Shor and Steane syndrome extraction and illustrate how it recovers a known configurable scheme for a CSS code.}

 The procedure with ${\mathcal{S}^{(M)}}=\mathcal{S}$ is exactly generalized Steane syndrome extraction. This reduces to standard Steane syndrome extraction for CSS codes and is equivalent to Knill syndrome extraction up to circuit identities \cite{GeneralSteane}.

Shor syndrome extraction is recovered when $\lvert{\mathcal{S}^{(M)}}\rvert=1$ and $\lvert{{\mathcal{S}}^{(LM)}}\rvert=0$, {where the \textit{cardinality}, denoted by $\lvert{\dots}\rvert$, gives the number of elements in the set, or equivalently here, the number of rows in the corresponding matrix representation. Note that if the stabilizer in $\mathcal{S}^{(M)}$ contains Pauli-$Y$s, the version of our procedure that uses controlled-$Y$ gates is required.}

\begin{proof}
{
    Per Step 3 of the procedure and $\lvert{\mathcal{S}^{(M)}}\rvert=1$, only one stabilizer of the ancilla state supports Pauli-$X$s; the rest are $Z$-type. In fact, as shown in Appendix \ref{shorsyndextrresult}, if the stabilizer being measured has weight $w$, the one ancilla stabilizer supporting Pauli-$X$s must be $X_1X_2 \dots X_w$ and the ancilla must have $w$ qubits.

   {Recall that this ancilla state encodes} $k=0$ logical qubits, so it has $w$ stabilizer generators. Recall also that in this case, all other $w-1$ stabilizer generators of the ancilla are $Z$-type and have a $+1$ phase. Therefore, the remaining stabilizers must be generated by $Z_{i}Z_{i+1} \forall i \in \{1,2,\dots, w-1\}$ to form a valid stabilizer state, per Appendix \ref{LemmaFormForStabs}. Recognize that this ancilla state is exactly the cat state used in Shor extraction.
   }

    Finally, the gates that remain in the circuit of our procedure must also be those of Shor syndrome extraction. This follows from a proof by contradiction, supposing that at least one of the gates differs from its counterpart in Shor syndrome extraction, and recalling that our procedure also uses $w$ controlled-Pauli gates, transversal with respect to the ancilla, to propagate the single weight-$w$ stabilizer in $\mathcal{S}^{(M)}$ onto the data.

    Thus, in this case, the ancilla and circuit of our procedure match that of Shor syndrome extraction.
\end{proof}

Note that Shor syndrome extraction cannot always be recovered using our procedure: while we can always set $\lvert{\mathcal{S}^{(M)}}\rvert=1$, it is not guaranteed that $\lvert{{\mathcal{S}}^{(LM)}}\rvert=0$. The algorithm in Appendix \ref{stabilizerform} gives a sufficient condition for $\lvert{{\mathcal{S}}^{(LM)}}\rvert=0$: each row of $\mathcal{S}^{(L)}$ has a unique column with a 1 that is also a zero column of $\mathcal{S}^{(M)}$. Regardless, even in cases where this condition is not satisfied, our procedure ends up measuring additional syndrome bits at no extra qubit cost, up to preparing the ancilla state and potentially needing to decompose Pauli-$Y$s measured on one qubit into a Pauli-$X$ and Pauli-$Z$ measured on two separate qubits.

{
Beyond these limiting cases, our procedure also exhibits a connection to known configurable syndrome-extraction schemes for CSS codes. Specifically, applying it to \textit{Steane's} $\llbracket 7,1,3\rrbracket$ \textit{code} reproduces an instance of Scheme A in \cite{Shinlin1}.}

\subsection{Fault-tolerance} \label{faultTolerancesubsec}

{
Faults in our scheme can introduce data errors through two mechanisms: direct error propagation and incorrect corrections prompted by an \textit{inaccurate} syndrome. Because the data-ancilla interactions are transversal, a single fault during these interactions can directly introduce an error on at most one data qubit. At the same time, faults that flip syndrome bits can cause a subsequent correction to introduce higher-weight data errors.

One may notice that these worst-case effects are the same as those encountered in Shor syndrome extraction, and that the same fault-tolerance argument used in Shor syndrome extraction \cite{Tansuwannont_2023,DiVincenzoShor,FTSErev} also applies to our scheme. Specifically, our procedure can achieve fault tolerance for a code correcting up to $t$ errors provided that, in the presence of up to $t$ total faults, the prepared ancilla contains \textit{sufficiently} few errors and the syndrome used for correction is \textit{accurate}.

Preparing stabilizer states with sufficiently few errors is a standalone problem, and how few errors are sufficient depends on the rest of the protocol. For example, fault-tolerant ancilla preparation is sufficient, but not always necessary: Pauli-$Z$ faults on the ancilla do not propagate to the data and their effects on the syndrome can be handled by the redundant measurements discussed below. Reducing this redundancy may, however, require stronger guarantees on these Pauli-$Z$ ancilla errors. Approaches to preparing ancilla states with sufficiently few errors include projecting onto the desired stabilizer eigenspaces using ancilla states that are easier to prepare with few errors \cite{GottesmanConstant,Steane97,khesin2026spidercatoptimalfaulttolerantcat}, \textit{distillation} \cite{Lai_2017egre,Zheng_2018erger}, and \textit{flag qubits} \cite{Prabhu_2023,Peham_2025}. 
}

Regarding syndrome accuracy, a sufficient condition, and the condition adopted here, is that the syndrome is not \textit{misdiagnosed}, meaning it is consistent with the errors present on the data immediately before its measurement. Like in Shor syndrome extraction, $t$ faults can cause a misdiagnosed syndrome in at most $t$ complete syndrome measurements. Therefore, an accurate syndrome can be guaranteed, up to $t$ faults, by repeating the procedure until $t+1$ consecutive complete syndrome measurements reveal the same syndrome. If we are to protect against at most a total of $t$ faults, this procedure will, at worst, have the syndrome change on every $(t+1)$-th complete syndrome measurement, and that can happen $t$ times, meaning $(t+1)^2$ complete syndrome measurements are needed in the worst case. Note however that this is a crude upper bound, and, in certain cases, our scheme can require fewer redundant syndrome measurements than Shor syndrome extraction, as elaborated on in the Discussion.

\subsection{Overhead and performance} \label{subsectionOverhead}

\begin{table}[b!]
\centering
\begin{tabular}{ |c||c|c|c| }
 \hline
 & Shor's & Ours & Steane's \\
 \hline \hline
 Peak ancilla width & $w$  & $W$ & $2n$\\
 \hline
 Total interaction depth & $mR(t)$ & $2 M R'(t)$ & $2$  \\
 \hline
\end{tabular}
 \caption{Summary of the worst-case overhead of Shor, Steane, and our syndrome extraction techniques. All variables are defined in the text. Note that the cost of preparing the different ancilla states is not considered.} 
\label{table:Tcr}
 \end{table}

To quantify the overhead, consider a stabilizer code that corrects all errors of weight up to $t$, using $n$ physical qubits and $m\leq n$ stabilizer generators, {each of weight at most $w\leq n$.} Let $1\leq M\leq m$ denote the number of {sets} used to \textit{cover} the stabilizer group. {That is, if $\mathcal{S}$ denotes a generating set for the code’s stabilizer group and  $\mathcal{S}_j\subseteq\langle\mathcal{S}\rangle$ is a generating set for the group of operators measured at time $j$, then $\langle\mathcal{S}\rangle=\langle\mathcal{S}_1\cup\mathcal{S}_2\cup\dots\cup\mathcal{S}_M\rangle$.

Let $W$, satisfying $w\leq W\leq 2n$, be the maximum \textit{combined weight} across all measurement generating sets $\mathcal{S}_j$.} The combined weight of stabilizer subset $\mathcal{S}_j$ in some sense quantifies the level of \textit{overlap} between the stabilizers in $\mathcal{S}_j$ and is calculated qubit-by-qubit across all the stabilizers in $\mathcal{S}_j$: if all operators in $\mathcal{S}_j$ have an $I$ on a qubit, assign that qubit 0; if all operators act with only identities and one type of Pauli ($X$s or $Y$s or $Z$s) on a qubit, assign it 1; otherwise assign it 2; the combined weight is the sum of these values across all $n$ qubits.

Let the functions $R(t)$ and $R'(t)$ denote the number of redundant syndrome measurements needed to guarantee fault tolerance in Shor syndrome extraction and our procedure, respectively. As discussed in Subsection \ref{faultTolerancesubsec}, $R(t), R'(t)\leq (t+1)^2$. {Recall that this upper bound is not necessarily tight.}

\begin{figure}[b!]
    \centering
    \includegraphics[width=\linewidth]{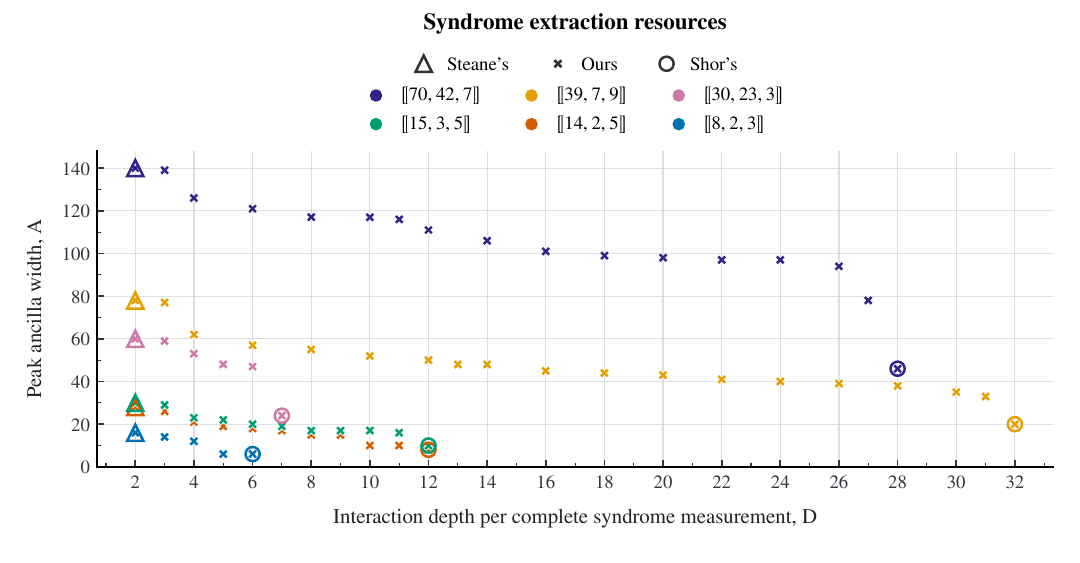}
    \caption{Costs of Shor, Steane, and our syndrome extraction circuits for the codes described in Appendix \ref{sims_appendix}. As an example, our procedure can measure the complete syndrome of the $\llbracket8,2,3\rrbracket$ code with a depth-$4$ circuit, up to the final single-qubit measurements, using at most $12$ ancilla qubits at any time. All costs exclude ancilla preparation and redundant syndrome measurements, as both are subject to optimizations beyond the scope of this work.}
    \label{fig:comparisontableresources}
\end{figure}

\begin{figure}[]
    \centering
    \includegraphics[width=\linewidth]{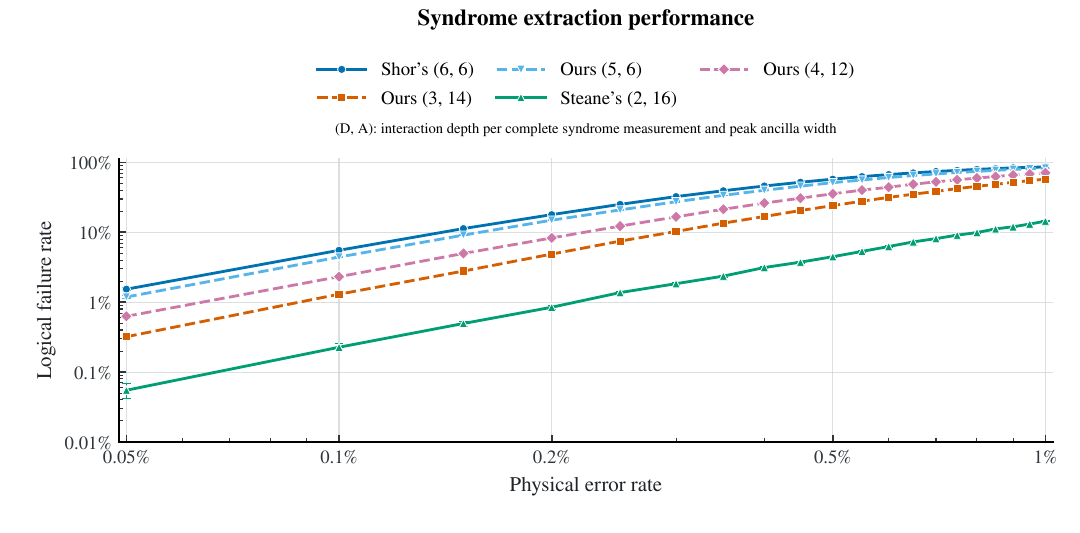}
    \caption{Performance of Shor, Steane, and our syndrome extraction circuits for the $\llbracket8,2,3\rrbracket$ code. Each curve corresponds to one of the five (D, A) points for this code in Figure \ref{fig:comparisontableresources}. The $x$-axis shows the physical error rate, and the $y$-axis shows the proportion of Monte Carlo runs in which the error remaining after the procedure was uncorrectable. Error bars show 95\% confidence intervals, and lines connect the data points to guide the eye. {The simulations use {circuit-level depolarizing noise and measurement errors}, with {ideal state preparation followed by independent single-qubit depolarizing noise} on each prepared qubit;} further simulation details are provided in Appendix \ref{sims_appendix}.}
    \label{fig:unbiased}
\end{figure}

The number of sequential layers of data–ancilla gates and the peak ancilla width of Shor, Steane, and our syndrome extraction techniques are summarized in terms of these parameters in Table \ref{table:Tcr}. Notice that because there is some freedom in the choice of $M$ and $W$, our procedure allows one to trade depth and qubit count, with Shor and Steane syndrome extraction often being opposite limits of this trade.

{
More concretely, Figure \ref{fig:comparisontableresources} compares peak ancilla width against the number of sequential layers of data–ancilla gates required for a complete syndrome measurement on six codes, using Shor’s scheme, Steane’s scheme, and several variants of our approach. For the four smaller codes, each plotted width is the exact minimum over all partitions of the generators. For the two larger codes, the widths are the best found by a \textit{random search}.
}

As for performance, Figures \ref{fig:unbiased} and \ref{fig:suppdata} compare the logical failure rates of our scheme with Shor- and Steane-style syndrome extraction across these four smaller codes. In these simulations, our scheme performs between Shor and Steane syndrome extraction. This is consistent with expectations: the number of ancilla qubits and gates our scheme requires lies between that of Shor and Steane, and thus so does the number of potential fault locations.

\section{Discussion} \label{Discussion}

{
Until quantum computers are built at substantially larger scales and with lower physical error rates, they will likely operate under tight resource constraints. There may therefore be regimes where one seeks to implement fault-tolerant quantum error correction but finds Shor syndrome extraction impractically deep and Steane syndrome extraction impractically wide. In this paper, we introduced a family of syndrome extraction gadgets for arbitrary stabilizer codes that interpolate between these extremes, enabling depth to be traded for width and vice versa. For a given code, this flexibility may help bring fault-tolerant syndrome extraction within a platform's resource limits.
}

Our work opens several directions for future research. Broadly, they fall into the following two categories: further generalizing the scheme, and reducing the scheme's cost in a fault-tolerant setting.

  {Here, we outline possible generalizations of the protocol.}

We have presented one method for constructing ancillas that are useful for syndrome extraction. {This construction, however, does not saturate the space of all ancilla states that enable transversal syndrome extraction; for example, it does not always recover Shor syndrome extraction. Characterizing a larger, ideally complete, space of ancilla states suitable for transversal syndrome extraction is therefore a natural direction for future work. A possible starting point is to revisit Step 2 of our procedure, which restricts the form of the ancilla stabilizers to simplify the proofs. Relaxing this assumption, and consequently taking more care in proving the necessary results, may lead to a larger ancilla space by allowing greater control over $\mathcal{S}^{(LL)}$. The main benefit of this extension may be a clearer connection to Shor syndrome extraction, since the stabilizers in $\mathcal{S}^{(LL)}$ have little to no effect on the protocol’s overhead.
}

The circuit itself offers another direction for generalization. Our construction already allows any transversal controlled-Pauli gate targeting the data, a hallmark of fault-tolerant circuits. Possible extensions {then include using circuit identities to modify gates and measurements, or introducing additional ancilla qubits and gates acting on them; the latter extension would substantially enlarge the search space, so some restrictions on circuit structure would likely still be necessary.
}

We now discuss potential improvements to our protocol aimed at optimizing its cost and performance in a fault-tolerant setting.

Our procedure inherits both strengths and weaknesses from Shor and Steane syndrome extraction. One major limitation, drawn from Shor syndrome extraction, is the need to verify the measured syndrome. In particular, a dominant contributor to the depth cost of our scheme is due to the factor $R'(t)\leq(t+1)^2$. Prior work has shown that the analogous factor in Shor syndrome extraction, $R(t)$, can be brought below $(t+1)^2$ \cite{Delfosse_2022,Tansuwannont_2023,ToddDSC,delfosse2020shortshorstylesyndromesequences}. In the following two paragraphs, we highlight some of the challenges and opportunities our procedure presents for reducing the {syndrome verification cost.}

{
A reason to suspect that our procedure might require fewer redundant syndrome measurements than Shor's, $R'(t)<R(t)$, is that our procedure potentially inherits some of the \textit{simultaneity} responsible for Steane syndrome extraction's exceptionally low depth. {That is, measuring sets of stabilizers simultaneously, rather than individually as in Shor syndrome extraction, can reduce the number of malignant fault locations between stabilizer measurements and thereby reduce the number of redundant syndrome measurements required \cite{delfosse2020shortshorstylesyndromesequences}.} We further expect that increasing the number of stabilizers measured simultaneously---by using fewer partitioning sets---may reduce the number of redundant syndrome measurements required. This expectation would be consistent with our procedure recovering Steane syndrome extraction, for which $R'(t)=1$ syndrome measurement suffices for fault tolerance.
}

{
Recall that we have so far considered only a naive strategy for redundant syndrome measurements in which each stabilizer measured is a fixed generator. This approach can be viewed as a special case of a more general approach: measuring carefully chosen products of stabilizers so that the syndrome is encoded in a form that also allows it to be checked for errors \cite{ToddDSC}. Notably, {this more general approach} can be adapted to our procedure to reduce the number of redundant syndrome measurements required. To maintain a low overhead, however, the redundant stabilizers must be covered by a small number of sets, each with relatively low combined weights. The challenge therefore is that these products of generators should not significantly increase the maximum weight of the operators being measured, as this would raise the lower bound of our procedure's ancilla width. While this problem may be difficult in general, restricting attention to specific code families may help.
}

Shifting focus away from minimizing the number of redundant syndrome measurements needed for fault tolerance, another direction is to {identify code families whose stabilizer structure is particularly well suited to our protocol}. Analogous work has been done for Shor syndrome extraction, where a landmark paper \cite{GottesmanConstant} showed that fault-tolerant quantum computing is achievable with asymptotically constant qubit overhead by leveraging \textit{high-rate}, \textit{quantum low-density parity-check} (QLDPC) codes \cite{GottesmanConstant, ExpanderCodes, Gottesman2022DC}. {Said protocol, however, comes with the drawbacks of high depth and low locality \cite{Terhal,Baspin2022,Gottesman2022DC}. Our scheme} may help mitigate these issues if one can identify codes whose stabilizers admit a cover {with a slowly growing combined weight $W$ and a slowly growing number of covering sets $M$}---that is, where our syndrome extraction circuit has slowly-growing width and depth. {Moreover, such codes may exhibit more favorable locality properties than QLDPC codes. }

For codes without an obviously favorable covering, it may be useful to devise an algorithm for finding favorable coverings---particularly for codes with many stabilizer generators. We observed, for the randomly chosen codes in Figure \ref{fig:comparisontableresources}, {many distinct partitions achieved optimal or near-optimal resource cost}. This suggests that an exhaustive search is likely unnecessary in most cases, and that simple heuristics may be sufficient to find favorable coverings efficiently.

{
Finally, one may also want to investigate the cost of preparing the ancilla states used in our procedure to better quantify the overhead \cite{GottesmanConstant,Steane97,Lai_2017egre,Zheng_2018erger,Prabhu_2023}. Also, new preparation methods should be explored, particularly those that suppress correlated $X$ and $Y$ errors; suppression of correlated $Z$ errors may be less critical, since the redundant syndrome measurements for syndrome verification can offset their effects.
}

\section*{Acknowledgments}

{
I am grateful to Todd Brun and Terry Farrelly for their helpful comments and suggestions on this work.

This work was supported in part by an Annenberg Fellowship from the University of Southern California and by NSF Grant FET-2316713.
}

\printbibliography

@article{Knill1998,
	title        = {Resilient Quantum Computation},
	author       = {Emanuel Knill  and Raymond Laflamme  and Wojciech Zurek},
	year         = 1998,
	journal      = {Science},
	volume       = 279,
	number       = 5349,
	pages        = {342--345},
	doi          = {10.1126/science.279.5349.342},
	url          = {https://www.science.org/doi/abs/10.1126/science.279.5349.342},
	eprint       = {https://www.science.org/doi/pdf/10.1126/science.279.5349.342}
}

@article{ThresholdTheoremConcat,
	title        = {Quantum Accuracy Threshold for Concatenated Distance-3 Codes},
	author       = {Aliferis, P. and Gottesman, D. and Preskill, J.},
	year         = 2006,
	journal      = {Quantum Information and Computation},
	publisher    = {Rinton Press, Incorporated},
	address      = {Paramus, NJ},
	volume       = 6,
	number       = 2,
	pages        = {97–165},
	issn         = {1533-7146},
	issue_date   = {March 2006},
	numpages     = 69
}

@inproceedings{Aharonov1997,
	title        = {Fault-Tolerant Quantum Computation with Constant Error},
	author       = {Aharonov, D.  and Ben-Or, M.},
	year         = 1997,
	booktitle    = {Proceedings of the Twenty-Ninth Annual ACM Symposium on Theory of Computing},
	venue        = {El Paso, Texas, USA},
	publisher    = {Association for Computing Machinery},
	location     = {New York, NY, USA},
	series       = {STOC '97},
	pages        = {176–188},
	doi          = {10.1145/258533.258579},
	isbn         = {0897918886},
	url          = {https://doi.org/10.1145/258533.258579},
	numpages     = 13
}

@article{GottesmanConstant,
	title        = {Fault-Tolerant Quantum Computation with Constant Overhead},
	author       = {Gottesman, Daniel},
	year         = 2014,
	journal      = {Quantum Information and Computation},
	publisher    = {Rinton Press, Incorporated},
	address      = {Paramus, NJ},
	volume       = 14,
	number       = {15–16},
	pages        = {1338–1372},
	issn         = {1533-7146},
	url          = {https://arxiv.org/abs/1310.2984},
	issue_date   = {November 2014},
	numpages     = 35
}

@article{ExpanderCodes,
	title        = {Constant Overhead Quantum Fault Tolerance with Quantum Expander Codes},
	author       = {Fawzi, O. and Grospellier, A. and Leverrier, A.},
	year         = 2020,
	journal      = {Communications of the ACM},
	publisher    = {Association for Computing Machinery},
	address      = {New York, NY, USA},
	volume       = 64,
	number       = 1,
	pages        = {106–114},
	doi          = {10.1145/3434163},
	issn         = {0001-0782},
	url          = {https://doi.org/10.1145/3434163},
	issue_date   = {January 2021},
	numpages     = 9
}

@inproceedings{Fawzi2022,
  author =	{Fawzi, Omar and M\"{u}ller-Hermes, Alexander and Shayeghi, Ala},
  title =	{{A Lower Bound on the Space Overhead of Fault-Tolerant Quantum Computation}},
  booktitle =	{13th Innovations in Theoretical Computer Science Conference (ITCS 2022)},
  pages =	{68:1--68:20},
  series =	{Leibniz International Proceedings in Informatics (LIPIcs)},
  ISBN =	{978-3-95977-217-4},
  ISSN =	{1868-8969},
  year =	{2022},
  volume =	{215},
  editor =	{Braverman, Mark},
  publisher =	{Schloss Dagstuhl -- Leibniz-Zentrum f{\"u}r Informatik},
  address =	{Dagstuhl, Germany},
  URL =		{https://drops.dagstuhl.de/entities/document/10.4230/LIPIcs.ITCS.2022.68},
  URN =		{urn:nbn:de:0030-drops-156649},
  doi =		{10.4230/LIPIcs.ITCS.2022.68}
}

@article{FTSErev,
	title        = {Syndromes and stabilizers: A review of fault-tolerant syndrome extraction in quantum error correction},
	author       = {Gorman, Miles D. and Farrelly, Terry},
	journal      = {(Manuscript in preparation)}
}

@inproceedings{Shor96,
	title        = {{Fault-Tolerant Quantum Computation}},
	author       = {Shor, Peter},
	year         = 1996,
	booktitle    = {Proceedings of 37th Conference on Foundations of Computer Science},
	publisher    = {IEEE Computer Society},
	pages        = {56--65},
	doi          = {10.1109/SFCS.1996.548464},
	url          = {https://doi.ieeecomputersociety.org/10.1109/SFCS.1996.548464}
}

@article{DiVincenzoShor,
	title        = {{Fault-Tolerant Error Correction with Efficient Quantum Codes}},
	author       = {David P. DiVincenzo and Peter W. Shor},
	year         = 1996,
	journal      = {Physical Review Letters},
	publisher    = {American Physical Society ({APS})},
	volume       = 77,
	number       = 15,
	pages        = {3260--3263},
	doi          = {10.1103/physrevlett.77.3260},
	url          = {https://doi.org/10.1103%2Fphysrevlett.77.3260}
}

@article{Steane97,
	title        = {{Active Stabilization, Quantum Computation, and Quantum State Synthesis}},
	author       = {Steane, Andrew M.},
	year         = 1997,
	journal      = {Phys. Rev. Lett.},
	publisher    = {American Physical Society},
	volume       = 78,
	pages        = {2252--2255},
	doi          = {10.1103/PhysRevLett.78.2252},
	url          = {https://link.aps.org/doi/10.1103/PhysRevLett.78.2252},
	issue        = 11,
	numpages     = {0}
}

@article{Knill1,
	title        = {{Scalable quantum computing in the presence of large detected-error rates}},
	author       = {Emanuel Knill},
	year         = 2005,
	journal      = {Phys. Rev. A},
	publisher    = {American Physical Society},
	volume       = 71,
	pages        = {042322},
	doi          = {10.1103/PhysRevA.71.042322},
	url          = {https://link.aps.org/doi/10.1103/PhysRevA.71.042322},
	issue        = 4,
	numpages     = 7
}

@article{Knill2,
	title        = {{Quantum computing with realistically noisy devices}},
	author       = {Emanuel Knill},
	year         = 2005,
	journal      = {Nature},
	publisher    = {Springer Science and Business Media {LLC}},
	volume       = 434,
	number       = 7029,
	pages        = {39--44},
	doi          = {10.1038/nature03350},
	url          = {https://doi.org/10.1038%2Fnature03350}
}

@article{Shinlin1,
	title        = {Between Shor and Steane: A Unifying Construction for Measuring Error Syndromes},
	author       = {Huang, Shilin and Brown, Kenneth R.},
	year         = 2021,
	month        = {8},
	journal      = {Phys. Rev. Lett.},
	publisher    = {American Physical Society},
	volume       = 127,
	pages        = {090505},
	doi          = {10.1103/PhysRevLett.127.090505},
	url          = {https://link.aps.org/doi/10.1103/PhysRevLett.127.090505},
	issue        = 9,
	numpages     = 5
}

@article{Shinlin2,
	title        = {Constructions for measuring error syndromes in Calderbank-Shor-Steane codes between Shor and Steane methods},
	author       = {Huang, Shilin and Brown, Kenneth R.},
	year         = 2021,
	month        = {8},
	journal      = {Phys. Rev. A},
	publisher    = {American Physical Society},
	volume       = 104,
	pages        = {022429},
	doi          = {10.1103/PhysRevA.104.022429},
	url          = {https://link.aps.org/doi/10.1103/PhysRevA.104.022429},
	issue        = 2,
	numpages     = 16
}

@article{ShinlinandYoder,
	title        = {Homomorphic Logical Measurements},
	author       = {Huang, Shilin and Jochym-O'Connor, Tomas and Yoder, Theodore J.},
	year         = 2023,
	month        = {7},
	journal      = {PRX Quantum},
	publisher    = {American Physical Society},
	volume       = 4,
	pages        = {030301},
	doi          = {10.1103/PRXQuantum.4.030301},
	url          = {https://link.aps.org/doi/10.1103/PRXQuantum.4.030301},
	issue        = 3,
	numpages     = 14
}

@article{Roffe,
	title        = {{Quantum error correction: an introductory guide}},
	author       = {Joschka Roffe},
	year         = 2019,
	journal      = {Contemporary Physics},
	publisher    = {Taylor & Francis},
	volume       = 60,
	number       = 3,
	pages        = {226--245},
	doi          = {10.1080/00107514.2019.1667078},
	url          = {https://doi.org/10.1080/00107514.2019.1667078},
	eprint       = {https://doi.org/10.1080/00107514.2019.1667078}
}

@book{nielsen,
	title        = {{Quantum Computation and Quantum Information}},
	author       = {Nielsen, Michael A. and Chuang, Isaac L.},
	year         = 2000,
	publisher    = {Cambridge University Press}
}

@misc{gottesmanStabsPaper,
	title        = {The Heisenberg Representation of Quantum Computers},
	author       = {Daniel Gottesman},
	year         = 1998,
	url          = {https://arxiv.org/abs/quant-ph/9807006},
	eprint       = {quant-ph/9807006},
	archiveprefix = {arXiv},
	primaryclass = {quant-ph}
}

@article{gottesman1997stabilizer,
	title        = {Stabilizer Codes and Quantum Error Correction},
	author       = {D. Gottesman},
	year         = 1997,
	journal      = {Caltech (PhD Thesis)}
}

@article{GeneralSteane,
  title = {Error rates and resource overheads of encoded three-qubit gates},
  author = {Takagi, Ryuji and Yoder, Theodore J. and Chuang, Isaac L.},
  journal = {Phys. Rev. A},
  volume = {96},
  issue = {4},
  pages = {042302},
  numpages = {13},
  year = {2017},
  month = {10},
  publisher = {American Physical Society},
  doi = {10.1103/PhysRevA.96.042302},
  url = {https://link.aps.org/doi/10.1103/PhysRevA.96.042302}
}

@article{Tansuwannont_2023,
	title        = {Adaptive syndrome measurements for Shor-style error correction},
	author       = {Tansuwannont, Theerapat and Pato, Balint and Brown, Kenneth R.},
	year         = 2023,
	month        = aug,
	journal      = {Quantum},
	publisher    = {Verein zur Forderung des Open Access Publizierens in den Quantenwissenschaften},
	volume       = 7,
	pages        = 1075,
	doi          = {10.22331/q-2023-08-08-1075},
	issn         = {2521-327X},
	url          = {http://dx.doi.org/10.22331/q-2023-08-08-1075}
}

@article{ToddDSC,
    title = "Quantum Data-Syndrome Codes",
    author = {Alexei Ashikhmin and Ching Yi Lai and Todd A. Brun},
    year = {2020},
    month = {3},
    doi = {10.1109/JSAC.2020.2968997},
    volume = {38},
    number = {3},
    pages = {449--462},
    journal = {IEEE Journal on Selected Areas in Communications},
    publisher = {IEEE}
}

@misc{delfosse2020shortshorstylesyndromesequences,
	title        = {Short Shor-style syndrome sequences},
	author       = {Nicolas Delfosse and Ben W. Reichardt},
	year         = 2020,
	url          = {https://arxiv.org/abs/2008.05051},
	eprint       = {2008.05051},
	archiveprefix = {arXiv},
	primaryclass = {quant-ph}
}

@article{Delfosse_2022,
	title        = {Beyond Single-Shot Fault-Tolerant Quantum Error Correction},
	author       = {Delfosse, Nicolas and Reichardt, Ben W. and Svore, Krysta M.},
	year         = 2022,
	month        = jan,
	journal      = {IEEE Transactions on Information Theory},
	publisher    = {Institute of Electrical and Electronics Engineers (IEEE)},
	volume       = 68,
	number       = 1,
	pages        = {287–301},
	doi          = {10.1109/tit.2021.3120685},
	issn         = {1557-9654},
	url          = {http://dx.doi.org/10.1109/TIT.2021.3120685}
}

@article{Terhal,
    title = {Tradeoffs for Reliable Quantum Information Storage in 2D Systems},
    author = {Bravyi, Sergey and Poulin, David and Terhal, Barbara},
    journal = {Phys. Rev. Lett.},
    volume = {104},
    issue = {5},
    pages = {050503},
    numpages = {4},
    year = {2010},
    month = {2},
    publisher = {American Physical Society},
    doi = {10.1103/PhysRevLett.104.050503},
    url = {https://link.aps.org/doi/10.1103/PhysRevLett.104.050503}
}

@article{Baspin2022,
    title = {Quantifying Nonlocality: How Outperforming Local Quantum Codes Is Expensive},
    author = {Baspin, Nou{\'e}dyn and Krishna, Anirudh},
    journal = {Phys. Rev. Lett.},
    volume = {129},
    issue = {5},
    pages = {050505},
    numpages = {5},
    year = {2022},
    month = {7},
    publisher = {American Physical Society},
    doi = {10.1103/PhysRevLett.129.050505},
    url = {https://link.aps.org/doi/10.1103/PhysRevLett.129.050505}
}

@article{Gottesman2022DC,
	title        = {Opportunities and Challenges in Fault-Tolerant Quantum Computation},
	author       = {Gottesman, D.},
	year         = 2022,
	eprint       = {2210.15844},
	eprinttype   = {arXiv},
	url          = {https://arxiv.org/abs/2210.15844},
	copyright    = {arXiv.org perpetual, non-exclusive license}
}

@book{Lidar_Brun_2013,
	title        = {Quantum Error Correction},
	editor       = {D. A. Lidar and T. A. Brun},
	year         = 2013,
	publisher    = {Cambridge University Press},
	place        = {Cambridge}
}

@article{Lai_2017egre,
    title = {Fault-tolerant preparation of stabilizer states for quantum Calderbank-Shor-Steane codes by classical error-correcting codes},
    author = {Lai, Ching-Yi and Zheng, Yi-Cong and Brun, Todd A.},
    journal = {Phys. Rev. A},
    volume = {95},
    issue = {3},
    pages = {032339},
    numpages = {11},
    year = {2017},
    month = {3},
    publisher = {American Physical Society},
    doi = {10.1103/PhysRevA.95.032339},
    url = {https://link.aps.org/doi/10.1103/PhysRevA.95.032339}
}

@article{Zheng_2018erger,
    title = {Efficient preparation of large-block-code ancilla states for fault-tolerant quantum computation},
    author = {Zheng, Yi-Cong and Lai, Ching-Yi and Brun, Todd A.},
    journal = {Phys. Rev. A},
    volume = {97},
    issue = {3},
    pages = {032331},
    numpages = {24},
    year = {2018},
    month = {3},
    publisher = {American Physical Society},
    doi = {10.1103/PhysRevA.97.032331},
    url = {https://link.aps.org/doi/10.1103/PhysRevA.97.032331}
}

@article{Prabhu_2023,
	title        = {Fault-tolerant syndrome extraction and cat state preparation with fewer qubits},
	author       = {Prabhu, Prithviraj and Reichardt, Ben W.},
	year         = 2023,
	month        = oct,
	journal      = {Quantum},
	publisher    = {Verein zur Forderung des Open Access Publizierens in den Quantenwissenschaften},
	volume       = 7,
	pages        = 1154,
	doi          = {10.22331/q-2023-10-24-1154},
	issn         = {2521-327X},
	url          = {http://dx.doi.org/10.22331/q-2023-10-24-1154}
}

@misc{codetables,
	title        = {{Bounds on the minimum distance of linear codes and quantum codes}},
	author       = {Grassl, Markus},
	year         = 2007,
	howpublished = {Online available at \url{http://www.codetables.de}}
}

@article{Stim,
	title        = {Stim: a fast stabilizer circuit simulator},
	author       = {Gidney, Craig},
	year         = 2021,
	month        = 7,
	journal      = {{Quantum}},
	publisher    = {{Verein zur F{\"{o}}rderung des Open Access Publizierens in den Quantenwissenschaften}},
	volume       = 5,
	pages        = 497,
	doi          = {10.22331/q-2021-07-06-497},
	issn         = {2521-327X},
	url          = {https://doi.org/10.22331/q-2021-07-06-497}
}

@misc{EncodingPackage,
	title        = {Ancilla-Free-Preparation-of-Stabilizer-States},
	author       = {Gorman, Miles D. and Farrelly, Terry},
	year         = 2025,
	journal      = {GitHub repository},
	publisher    = {GitHub},
	howpublished = {\url{https://github.com/MilesGorman/Ancilla-Free-Preparation-of-Stabilizer-States}}
}

@book{gottesman2024surviving,
  author    = {Daniel Gottesman},
  title     = {Surviving as a Quantum Computer in a Classical World},
  year      = {2024},
  note      = {Lecture notes / draft textbook available at \url{https://www.cs.umd.edu/~dgottesm/}},
  publisher = {University of Maryland}
}

@misc{gottesman2009introductionquantumerrorcorrection,
      title={An Introduction to Quantum Error Correction and Fault-Tolerant Quantum Computation}, 
      author={Daniel Gottesman},
      year={2009},
      eprint={0904.2557},
      archivePrefix={arXiv},
      primaryClass={quant-ph},
      url={https://arxiv.org/abs/0904.2557}, 
}

@misc{khesin2026spidercatoptimalfaulttolerantcat,
      title={SpiderCat: Optimal Fault-Tolerant Cat State Preparation}, 
      author={Andrey Boris Khesin and Sarah Meng Li and Boldizsár Poór and Benjamin Rodatz and John van de Wetering and Richie Yeung},
      year={2026},
      eprint={2603.05391},
      archivePrefix={arXiv},
      primaryClass={quant-ph},
      url={https://arxiv.org/abs/2603.05391}, 
}

@article{Peham_2025,
   title={Automated Synthesis of Fault-Tolerant State Preparation Circuits for Quantum Error-Correction Codes},
   volume={6},
   ISSN={2691-3399},
   pages = {020330},  
   url={http://dx.doi.org/10.1103/PRXQuantum.6.020330},
   DOI={10.1103/prxquantum.6.020330},
   number={2},
   journal={PRX Quantum},
   publisher={American Physical Society (APS)},
   author={Peham, Tom and Schmid, Ludwig and Berent, Lucas and Müller, Markus and Wille, Robert},
   year={2025},
   month=May }

\begin{appendices}

\newpage

\section{Algorithm for partitioning
  \texorpdfstring{$\langle \mathcal{S}^{(L)} \rangle$}{<S(L)>}}
\label{stabilizerform}

    Given a set of stabilizer generators $\mathcal{S}^{(M)}$ to be measured in the current round, and a set of stabilizer generators $\mathcal{S}^{(L)}$ to be measured in a later round, our procedure requires rewriting the generating set $\mathcal{S}^{(L)}$ as

    \begin{equation}
        \mathcal{S}^{(LM)} \cup \mathcal{S}^{(LL)}
        \label{transformationrule}
    \end{equation}
    
    \noindent such that:
    \begin{enumerate}

        \item Every zero column of $\mathcal{S}^{(M)}$ is also a zero column of $\mathcal{S}^{(LM)}$.
        \item Every row of ${\mathcal{S}}^{(LL)}$ has an entry of 1 in some column that no other row of $\mathcal{S}^{(M)}\cup{\mathcal{S}}^{(LL)}$ has a 1 in.
        \item The group generated remains the same: $\langle{\mathcal{S}}^{(LM)} \cup {\mathcal{S}}^{(LL)}\rangle = \langle {\mathcal{S}}^{(L)}\rangle$.
    \end{enumerate}

    \noindent Below is an algorithm for achieving such a form for any stabilizer code.

    \begin{enumerate}[label*=\arabic*.]
        \item Initialize ${\mathcal{S}}^{(LM)}\leftarrow \mathcal{S}^{(L)}$, ${\mathcal{S}}^{(LL)} \leftarrow\emptyset$, ${\mathcal{U}}\leftarrow \emptyset$, and let $\mathcal{C}^{(M)}$ denote the indices of zero columns in $\mathcal{S}^{(M)}$.
        \item  For each index $c\in\mathcal{C}^{(M)}$:
        \begin{enumerate}[label*=\arabic*.]
            \item If all rows of ${\mathcal{S}}^{(LM)}$ have an entry of 0 in column $c$, continue.
            \item Else:
            \begin{enumerate}[label*=\arabic*.]
                \item Append $c$ to ${\mathcal{U}}$.
                \item Select a row $r$ from ${\mathcal{S}}^{(LM)}$ with a 1 in column $c$. 
                \item Remove $r$ from ${\mathcal{S}}^{(LM)}$ and append it to ${\mathcal{S}}^{(LL)}$.
                \item For every other row $t$ in ${\mathcal{S}}^{(LM)} \cup {\mathcal{S}}^{(LL)}$ that has a 1 in column $c$, replace $t \leftarrow t \oplus r$.
            \end{enumerate}
        \end{enumerate}
    \end{enumerate}

    Due to steps 2.1, 2.2.3, and 2.2.4, by the end of the procedure, all rows remaining in ${\mathcal{S}}^{(LM)}$ must have an entry of 0 in columns indexed by $\mathcal{C}^{(M)}$, i.e., in the zero columns of $\mathcal{S}^{(M)}$. Therefore, ${\mathcal{S}}^{(LM)}$ indeed has the form described by the first property under Eq.\ \ref{transformationrule}. 
    
    Meanwhile, per step 2.2.3, each row of ${\mathcal{S}}^{(LL)}$ contains an entry of 1---in a column indexed by some $c\in{\mathcal{U}} \subseteq \mathcal{C}^{(M)}$---that, due to step 2.2.4, no other row of ${\mathcal{S}}^{(LL)}$ has a 1 in, fulfilling the second property under Eq.\ \ref{transformationrule}. 

    The third property under Eq.\ \ref{transformationrule} is indeed satisfied since ${\mathcal{S}}^{(LM)} \cup {\mathcal{S}}^{(LL)}$ and ${\mathcal{S}}^{(L)}$ have identical \textit{row spaces}.

Note that a simpler, sufficient condition is that each row of $\mathcal{S}^{(L)}$ has a 1 in a unique zero column of $\mathcal{S}^{(M)}$. This can effectively be achieved for any stabilizer code by first putting the code's stabilizer matrix $\mathcal{S}$ into \textit{reduced row-echelon form} and then choosing the partitioning sets $\mathcal{S}^{(M)}$ and $\mathcal{S}^{(L)}$. In such a case, ${\mathcal{S}}^{(LM)}=\emptyset$, ${\mathcal{S}}^{(LL)}=\mathcal{S}^{(L)}$. However, row reduction generally changes the generators, so this approach does not allow an arbitrary, previously chosen set $\mathcal{S}^{(M)}$ to be measured directly, whereas our algorithm does.

\newpage
\section{Pedagogical example} \label{secexample}

In this section, we demonstrate our procedure on the five-qubit code, whose stabilizer generators are defined in the text surrounding Eq.\ \ref{fivequbitcodestabs}.
Specifically, we extract the syndrome bits of the stabilizers $X_1Z_2Z_3X_4$ and $X_2Z_3Z_4X_5$ simultaneously. The steps of the procedure are as follows:

\begin{enumerate}
    \item[1.] Choose $\mathcal{S}^{(M)}=\{X_1Z_2Z_3X_4,X_2Z_3Z_4X_5\}$, so
    $\mathcal{S}^{(L)}=\{X_1 X_3 Z_4 Z_5, Z_1 X_2 X_4 Z_5 \}$. 
    
    \item[2.] This partition satisfies the required form, where ${\mathcal{S}}^{(LL)}=\{X_1 X_3 Z_4 Z_5, Z_1 X_2 X_4 Z_5 \}$ and ${\mathcal{S}}^{(LM)}=\emptyset$. 

    \item[3, 4.] The ancilla is, tentatively, to have the stabilizer generators

        \begin{equation}
       \left( \begin{array}{@{}  c |  c @{}}
       \begin{matrix}
        1 & 0 & 0 & 1 & 0 & 0 & 1 & 1 & 0 & 0 \\
        0 & 1 & 0 & 0 & 1 & 0 & 0 & 1 & 1 & 0 \\
        1 & 0 & 1 & 0 & 0 & 0 & 0 & 0 & 1 & 1 \\
        0 & 1 & 0 & 1 & 0 & 1 & 0 & 0 & 0 & 1 \\

        0 & 0 & 0 & 0 & 0 & 0 & 0 & 0 & 0 & 0 \\
        0 & 0 & 0 & 0 & 0 & 0 & 0 & 0 & 0 & 0 \\
        0 & 0 & 0 & 0 & 0 & 0 & 0 & 0 & 0 & 0 \\
        0 & 0 & 0 & 0 & 0 & 0 & 0 & 0 & 0 & 0 \\

        0 & 0 & 0 & 0 & 0 & 0 & 0 & 0 & 0 & 0 \\
        0 & 0 & 0 & 0 & 0 & 0 & 0 & 0 & 0 & 0
       \end{matrix}
       &
       \begin{matrix}
        0 & 0 & 0 & 0 & 0 & 1 & 0 & 0 & 1 & 0 \\
        0 & 0 & 0 & 0 & 0 & 0 & 1 & 0 & 0 & 1 \\
        0 & 0 & 0 & 0 & 0 & 1 & 0 & 1 & 0 & 0 \\
        0 & 0 & 0 & 0 & 0 & 0 & 1 & 0 & 1 & 0 \\

        0 & 1 & 1 & 0 & 0 & 1 & 0 & 0 & 1 & 0 \\
        0 & 0 & 1 & 1 & 0 & 0 & 1 & 0 & 0 & 1 \\
        0 & 0 & 0 & 1 & 1 & 1 & 0 & 1 & 0 & 0 \\
        1 & 0 & 0 & 0 & 1 & 0 & 1 & 0 & 1 & 0 \\

        1 & 1 & 1 & 1 & 1 & 0 & 0 & 0 & 0 & 0 \\
        0 & 0 & 0 & 0 & 0 & 1 & 1 & 1 & 1 & 1
       \end{matrix}
       \end{array} \right).
   \end{equation}

 \noindent Rows 3 and 4, and columns 3, 6, 10, 13, 16, and 20 are removed. The ancilla's stabilizer generators are then

    \begin{equation}
       \left( \begin{array}{@{}  c |  c @{}}
       \begin{matrix}
        1 & 0  & 1 & 0  & 1 & 1 & 0  \\
        0 & 1  & 0 & 1  & 0 & 1 & 1  \\

        0 & 0  & 0 & 0  & 0 & 0 & 0  \\
        0 & 0  & 0 & 0  & 0 & 0 & 0  \\
        0 & 0  & 0 & 0  & 0 & 0 & 0  \\
        0 & 0  & 0 & 0  & 0 & 0 & 0  \\

        0 & 0  & 0 & 0  & 0 & 0 & 0  \\
        0 & 0  & 0 & 0  & 0 & 0 & 0
       \end{matrix}
       &
       \begin{matrix}
        0 & 0  & 0 & 0  & 0 & 0 & 1  \\
        0 & 0  & 0 & 0  & 1 & 0 & 0  \\

        0 & 1  & 0 & 0  & 0 & 0 & 1  \\
        0 & 0  & 1 & 0  & 1 & 0 & 0  \\
        0 & 0  & 1 & 1  & 0 & 1 & 0  \\
        1 & 0  & 0 & 1  & 1 & 0 & 1  \\

        1 & 1  & 1 & 1  & 0 & 0 & 0  \\
        0 & 0  & 0 & 0  & 1 & 1 & 1
       \end{matrix}
       \end{array} \right).
   \end{equation}

     \noindent Notice as well that row 7 is the linear combination of rows 3, 4, and 6, so it is redundant and can be trivially removed, giving the ancilla's stabilizer generators as

       \begin{equation}
       \left( \begin{array}{@{}  c |  c @{}}
       \begin{matrix}
        1 & 0  & 1 & 0  & 1 & 1 & 0  \\
        0 & 1  & 0 & 1  & 0 & 1 & 1  \\

        0 & 0  & 0 & 0  & 0 & 0 & 0  \\
        0 & 0  & 0 & 0  & 0 & 0 & 0  \\
        0 & 0  & 0 & 0  & 0 & 0 & 0  \\
        0 & 0  & 0 & 0  & 0 & 0 & 0  \\

        0 & 0  & 0 & 0  & 0 & 0 & 0
       \end{matrix}
       &
       \begin{matrix}
        0 & 0  & 0 & 0  & 0 & 0 & 1  \\
        0 & 0  & 0 & 0  & 1 & 0 & 0  \\

        0 & 1  & 0 & 0  & 0 & 0 & 1  \\
        0 & 0  & 1 & 0  & 1 & 0 & 0  \\
        0 & 0  & 1 & 1  & 0 & 1 & 0  \\
        1 & 0  & 0 & 1  & 1 & 0 & 1  \\

        0 & 0  & 0 & 0  & 1 & 1 & 1
       \end{matrix}
       \end{array} \right).
           \label{ExampleStabilizers}
   \end{equation}

 Apply the circuit shown in Figure \ref{ExampleCirc5Qubit}, where the ancilla $\ket{a}$ is prepared in the simultaneous $+1$ eigenstate of Eq.\ \ref{ExampleStabilizers}.

\item[5.] Let the circuit's measurement outcomes be $\left(m_1, m_2,\dots, m_7 \right)\in \mathbb{Z}_2^7$. The syndrome bits of $\mathcal{S}^{(M)}$ are

\begin{equation}
       \left( 
       \begin{matrix}
        1 & 0  & 1 & 0 & 1 & 1 & 0\\
        0 & 1  & 0 & 1 & 0 & 1 & 1
       \end{matrix}
 \right)
       \left(
       \begin{matrix}
        m_1\\ m_2 \\ \vdots \\ m_7
       \end{matrix}
        \right).
   \end{equation}

\item[6.] To obtain the complete syndrome, these steps can be repeated with, {say}, $\mathcal{S}^{(M)}$ and $\mathcal{S}^{(L)}$ swapped. 
\end{enumerate}

\begin{figure}[h!]
    \begin{center}
            \includegraphics{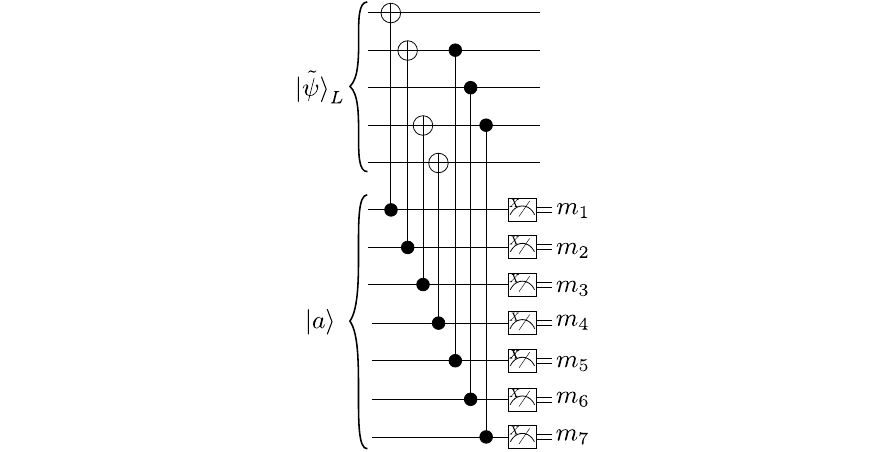}
    \end{center}
    \caption{
    Circuit implementing our syndrome extraction procedure for the five-qubit code, with $\mathcal{S}^{(M)}=\{X_1Z_2Z_3X_4,X_2Z_3Z_4X_5\}$. The input state $\ket{\Tilde{\psi}}_L$ is a potentially erroneous logical state, and $\ket{a}$ is an ancilla initialized {with the stabilizers in Eq.\ \ref{ExampleStabilizers}.}}
    \label{ExampleCirc5Qubit}
\end{figure}

\newpage
\section{The ancilla state is valid} \label{validstabs}

\begin{theorem}
Let $\mathcal{S}$ be a set of stabilizer generators for some code. According to steps 1 and 2 of the procedure, partition the stabilizer generators such that $\mathcal{S}=\mathcal{S}^{(M)}\cup \mathcal{S}^{(L)}$ and $\langle \mathcal{S}^{(L)}\rangle = \langle \mathcal{S}^{(LM)}\cup \mathcal{S}^{(LL)} \rangle$ where $\mathcal{S}^{(M)}$ is chosen arbitrarily; ${\mathcal{S}}^{(LM)}$ has zero columns in at least the same locations as $\mathcal{S}^{(M)}$; and every row of ${\mathcal{S}}^{(LL)}$ has an entry of 1 in a column that no other row of $\mathcal{S}^{(M)}\cup{\mathcal{S}}^{(LL)}$ has a 1 in. 

The ancilla constructed in steps 3 and 4 of the procedure---obtained by removing from the Steane syndrome extraction ancilla in Eq.\ \ref{AncillaStabsProcedure} all qubits corresponding to zero columns of $\mathcal{S}^{(M)}$ and all rows corresponding to $\mathcal{S}^{(LL)}$---is a valid stabilizer state. 
\end{theorem}
    
\begin{proof}
    Let $\mathcal{C}^{(M)}$ and ${\mathcal{C}^{(LM)}}$ denote the sets of indices corresponding to zero columns in $\mathcal{S}^{(M)}$ and ${\mathcal{S}}^{(LM)}$, respectively. 
   Also, define $\mathcal{U}$ as in Appendix \ref{stabilizerform}, 
   i.e., consisting of column indices in $\mathcal{S}^{(LL)}$ that contain a 1 in columns where no other row of $\mathcal{S}^{(M)}\cup{\mathcal{S}}^{(LM)}\cup{\mathcal{S}}^{(LL)}$ has a 1. 
   Recall from Appendix \ref{stabilizerform} and step 2 of the procedure that each row of ${\mathcal{S}}^{(LL)}$ contains a column indexed by $c\in\mathcal{U}\subseteq\mathcal{C}^{(M)}$ with a single 1, and $\lvert{\mathcal{U}}\rvert=\lvert{\mathcal{S}^{(LL)}}\rvert$. 

     Recalling Eq.\ \ref{AncillaStabsProcedure}, the initial parity-check matrix of the ancilla state is the one used in Steane syndrome extraction:

    \begin{equation}
       \left( \begin{array}{@{}c@{}|c@{}}
       \begin{matrix}
          {\mathcal{S}}_X^{(LL)} & {\mathcal{S}}_Z^{(LL)} \\
          {\mathcal{S}}_X^{(LM)} & {\mathcal{S}}_Z^{(LM)} \\
          \mathcal{S}_X^{(M)} & \mathcal{S}_Z^{(M)} \\
          0 & 0 \\
          0 & 0
       \end{matrix}
       &
       \begin{matrix}
          0 & {\mathcal{S}}_X^{(LL)} \\
          0 & {\mathcal{S}}_X^{(LM)} \\
          0 & \mathcal{S}_X^{(M)} \\
          \mathcal{S}_Z & \mathcal{S}_X \\
          \mathcal{L}_Z & \mathcal{L}_X
       \end{matrix}
       \end{array} \right)
   \end{equation}

   \noindent where the $X$ and $Z$ subscripts were defined in Eq.\ \ref{stabilzierdecomp}.
    This ancilla state utilizes $2n$ qubits and is stabilized by $2n$ independent generators. Removing the $\lvert{{\mathcal{S}}^{(LL)}}\rvert$ generators corresponding to $\mathcal{S}^{(LL)}$ gives a $2n$-qubit code with $2n-\lvert{{\mathcal{S}}^{(LL)}}\rvert$ generators, namely

           \begin{equation}
       \left( \begin{array}{@{}c@{}|c@{}}
       \begin{matrix}
          {\mathcal{S}}_X^{(LM)} & {\mathcal{S}}_Z^{(LM)} \\
           \mathcal{S}_X^{(M)} & \mathcal{S}_Z^{(M)} \\
          0 & 0 \\
          0 & 0
       \end{matrix}
       &
       \begin{matrix}
          0 & {\mathcal{S}}_X^{(LM)} \\
           0 & \mathcal{S}_X^{(M)} \\
          \mathcal{S}_Z & \mathcal{S}_X \\
          \mathcal{L}_Z & \mathcal{L}_X
       \end{matrix}
       \end{array} \right).
           \label{validancillastabmatrix}
   \end{equation}
    
    \noindent Note: although we intended to exclude the ancilla stabilizer generators associated with ${\mathcal{S}}^{(LM)}$ as well, we retain them to simplify the proof and because doing so does not increase the ancilla width, up to preparing the ancilla. This is because $\mathcal{C}^{(M)} \subseteq \mathcal{C}^{(LM)}$ and the qubits in $\mathcal{C}^{(M)}$ are later deleted.

    To restore the ancilla to a stabilizer state, an additional $\lvert{{\mathcal{S}}^{(LL)}}\rvert$ independent generators are needed. Doing this ensures that the ancilla state is fixed and guarantees that no additional Pauli-$X$s are present in the stabilizer group, which could lead to unintended measurements on the data.
    
    We will now show that appending ${\mathcal{Z}^{(LL)}}\equiv\{Z_\sigma | \sigma \in {\mathcal{U}}\}$ to the stabilizer generators of Eq.\ \ref{validancillastabmatrix} restores the ancilla to a stabilizer state. That is, consider an ancilla stabilized by

          \begin{equation}
       \left( \begin{array}{@{}c@{}|c@{}}
       \begin{matrix}
          0 & 0 \\
           {\mathcal{S}}_X^{(LM)} & {\mathcal{S}}_Z^{(LM)} \\
          \mathcal{S}_X^{(M)} & \mathcal{S}_Z^{(M)} \\
          0 & 0 \\
          0 & 0
       \end{matrix}
       &
       \begin{matrix}
         \multicolumn{2}{c}{{\mathcal{Z}}^{(LL)}}\\
          0 & {\mathcal{S}}_X^{(LM)} \\
          0 & \mathcal{S}_X^{(M)} \\
          \mathcal{S}_Z & \mathcal{S}_X \\
          \mathcal{L}_Z & \mathcal{L}_X
       \end{matrix}
       \end{array} \right).
       \label{anothervalidstabmatrix}
   \end{equation}

    To establish the independence of the proposed ancilla's stabilizer generators, recall that each row of ${{\mathcal{S}}^{(LL)}}$ has a unique column $c\in {\mathcal{U}}$ containing a single 1. Therefore, each of the $\lvert{{\mathcal{S}}^{(LL)}}\rvert$ operators $Z_{\sigma}\in {\mathcal{Z}^{(LL)}}$ anticommute with exactly one corresponding stabilizer generator $S_\sigma \in ( \   {\mathcal{S}}_X^{(LL)} \quad {\mathcal{S}}_Z^{(LL)} \quad | \quad 0 \quad {\mathcal{S}}_X^{(LL)} )$.
    Since $Z_\sigma$ anticommutes with $S_\sigma$, while all other operators in Eq.\ \ref{anothervalidstabmatrix} commute with $S_\sigma$, it follows that $Z_\sigma$
    cannot be expressed as a product of these other operators and so is independent from them.
    Because this argument holds for each of the $\lvert{{\mathcal{S}}^{(LL)}}\rvert$ operators in ${\mathcal{Z}^{(LL)}}$, appending ${\mathcal{Z}^{(LL)}}$ to the $2n-\lvert{{\mathcal{S}}^{(LL)}}\rvert$ generators in Eq.\ \ref{validancillastabmatrix} yields a set of $2n$ independent operators.

    To verify that the operators in the resulting set, Eq.\ \ref{anothervalidstabmatrix}, define a valid stabilizer group, we must confirm that they mutually commute. Recall the inclusion $ \mathcal{C}^{(M)} \subseteq {\mathcal{C}^{(LM)}}$, implying that the left half of the tentative parity check matrix in Eq.\ \ref{validancillastabmatrix} has zero columns indexed by $\mathcal{C}^{(M)}$.
    Because ${\mathcal{U}\subseteq \mathcal{C}^{(M)}}$, the operators in ${\mathcal{Z}^{(LL)}}$ commute with all the stabilizers in Eq.\ \ref{validancillastabmatrix}. Moreover, the operators in ${\mathcal{Z}^{(LL)}}$ are all $Z$-type, so they mutually commute. Therefore, the stabilizers in Eq.\ \ref{anothervalidstabmatrix} satisfy commutativity, and thus describe a valid stabilizer state.

    Finally, to see that removing the qubits indexed by $\mathcal{C}^{(M)}$ does not affect the ancilla's validity, observe, using the same reasoning as above, that actually all operators in $\mathcal{Z}^{(M)}\equiv\{Z_\sigma | \sigma \in {\mathcal{C}^{(M)}}\}$ commute with the proposed stabilizers in Eq.\ \ref{anothervalidstabmatrix}. Because all the operators in $\mathcal{Z}^{(M)}$ commute with the stabilizers of the proposed stabilizer state, they too must be in the state's stabilizer group, as argued in Lemma \ref{basicresultonstabs}, noting that the state's stabilizer generators from Eq.\ \ref{anothervalidstabmatrix} all have a $+1$ phase.
    Therefore, the ancilla's (over-complete) stabilizer generators include

       \begin{equation}
       \left( \begin{array}{@{}c@{}|c@{}}
       \begin{matrix}
          0 & 0 \\
          {\mathcal{S}}_X^{(LM)} & {\mathcal{S}}_Z^{(LM)} \\
          \mathcal{S}_X^{(M)} & \mathcal{S}_Z^{(M)} \\
          0 & 0 \\
          0 & 0
       \end{matrix}
       &
       \begin{matrix}
          \multicolumn{2}{c}{\mathcal{Z}^{(M)}}\\
           0 & {\mathcal{S}_X}^{(LM)} \\
          0 & \mathcal{S}_X^{(M)} \\
          \mathcal{S}_Z & \mathcal{S}_X \\
          \mathcal{L}_Z & \mathcal{L}_X
       \end{matrix}
       \end{array} \right)
          \label{ExplanationAncillaStabs}.
   \end{equation}

   \noindent Observe that after removing the $\lvert{\mathcal{Z}^{(M)}}\rvert$ stabilizers in $\mathcal{Z}^{(M)}$ and the $\lvert{\mathcal{Z}^{(M)}}\rvert$ qubits indexed by $\mathcal{C}^{(M)}$, the remaining stabilizers still define a valid stabilizer group. This is because we can, prior to removing qubits, choose an equivalent generating set in which no stabilizers besides ${\mathcal{Z}^{(M)}}$ have support on the qubits in $\mathcal{C}^{(M)}$. As a result, removing these qubits does not alter the commutation relations among the remaining stabilizers. With this modified description for the stabilizer generators, observe further that since we remove exactly as many generators as qubits, the ancilla retains a stabilizer state. Noting that each operator in $\mathcal{Z}^{(M)}$ is guaranteed to be a generator because the qubits they act on have been removed from the support of all other generators in the ancilla, completing the argument.
\end{proof}

\begin{lemma}
Let $\mathcal{S}$ be the stabilizer group of an $\llbracket n,k\rrbracket$ code and 

\begin{equation}
        \mathcal{C}(\mathcal{S})=\{P\in\mathcal{G}_n : PS=SP\forall S \in\mathcal{S}\}
\end{equation}

\noindent be the \textit{centralizer} of the stabilizer group $\mathcal{S}$.

If $k=0$, then $\mathcal{C}(\mathcal{S})=\{\pm1,\pm i\}\times\mathcal{S}$, i.e., any Pauli that commutes with all the stabilizer generators must itself lie in the stabilizer group $\mathcal{S}$, up to a phase. 
    \label{basicresultonstabs}
\end{lemma}

\begin{proof}
Because  $\mathcal{S}\subset\mathcal{G}_n$ and stabilizers mutually commute regardless of their phase, we have
$\{\pm1,\pm i\}\times\mathcal{S}\subseteq \mathcal{C}(\mathcal{S})$.

For the reverse inclusion, from \cite{Lidar_Brun_2013}, we know ${\mathcal{C}(\mathcal{S}) / \mathcal{S}}\cong {\mathcal{G}_{k}}$. In the case that $k=0$, this implies $\mathcal{C}(\mathcal{S}) / \mathcal{S}=\{\pm1,\pm i\}$. Equivalently, for every $P\in{\mathcal{C}(\mathcal{S})}$, the \textit{coset} $P\mathcal{S}= \phi \mathcal{S}$ for some $\phi \in \{\pm1,\pm i\}$. This means there exists an $S\in\mathcal{S}$ such that $P=\phi S$. Therefore, $\mathcal{C}(\mathcal{S})\subseteq \{\pm1,\pm i\}\times\mathcal{S}$.

Combining both inclusions, we have equality: $\mathcal{C}(\mathcal{S})=\{\pm1,\pm i\}\times\mathcal{S}$.
\end{proof}

\newpage
\section{The intended syndrome bits are extracted} \label{ProofSyndromeExtraction}

\begin{theorem}
Let $\mathcal{S}$ be a set of stabilizer generators for some code. The ancilla stabilizer state $\ket{a}$ of our procedure is given in Eq.\ \ref{ExplanationAncillaStabs}, where the stabilizers are partitioned such that $\mathcal{S}=\mathcal{S}^{(M)}\cup \mathcal{S}^{(L)}$ and $\langle \mathcal{S}^{(L)}\rangle = \langle \mathcal{S}^{(LM)}\cup \mathcal{S}^{(LL)} \rangle$ where $\mathcal{S}^{(M)}$ is chosen arbitrarily; ${\mathcal{S}}^{(LM)}$ has zero columns in at least the same locations as $\mathcal{S}^{(M)}$; every row of ${\mathcal{S}}^{(LL)}$ has an entry of 1 in a column that no other row of $\mathcal{S}^{(M)}\cup{\mathcal{S}}^{(LL)}$ has a 1 in; and the $X$ and $Z$ subscripts are defined in Eq.\ \ref{stabilzierdecomp}. Also, let $\mathcal{C}^{(M)}$ denote the indices of zero columns in $\mathcal{S}^{(M)}$ and $\mathcal{Z}^{(M)}\equiv\{Z_\sigma | \sigma \in {\mathcal{C}^{(M)}}\}$.

When this ancilla state $\ket{a}$ is used in the circuit depicted in Figure \ref{fig:ShorAndSteane}b, only the syndrome bits of the generators $\mathcal{S}^{(M)} \cup {\mathcal{S}}^{(LM)}$ are extracted. Moreover, the ancilla qubits indexed by elements of $\mathcal{C}^{(M)}$, along with any gates acting on them, can be deleted trivially.
\end{theorem}

\begin{proof}
 The stabilizers and logical operators of the data-ancilla system are

    \begin{equation}
       \left( \begin{array}{@{}c@{}|c@{}|c@{}}
       \begin{matrix}
           s\\0\\0\\0\\0\\0
       \end{matrix}
       &
       \begin{matrix}
           \mathcal{S}_X & 0 & 0\\
           0&0&0\\
           0 & {\mathcal{S}}_X^{(LM)} & {\mathcal{S}}_Z^{(LM)} \\
          0 & \mathcal{S}_X^{(M)} & \mathcal{S}_Z^{(M)} \\
          0 & 0 & 0 \\
          0 & 0 & 0 \\
       \end{matrix}
       &
       \begin{matrix}
          \mathcal{S}_Z & 0 & 0\\
         0 & \multicolumn{2}{c}{\mathcal{Z}^{(M)}}\\
          0 & 0 & {\mathcal{S}}_X^{(LM)} \\
          0 & 0 & \mathcal{S}_X^{(M)} \\
          0 & \mathcal{S}_Z & \mathcal{S}_X \\
          0 & \mathcal{L}_Z & \mathcal{L}_X \\
       \end{matrix}
          \\ \hline
          \begin{matrix}
          \phantom{0}
           \end{matrix}
           &
           \begin{matrix}
                          \textover[c]{$\mathcal{L}_X$}{$\mathcal{S}_X$} & \textover[c]{$0$}{$\mathcal{S}_X^{(LM)}$} & \textover[c]{$0$}{$\mathcal{S}_Z^{(LM)}$}\\
       \end{matrix}
       &
       \begin{matrix}
           \textover[c]{$\mathcal{L}_Z$}{$\mathcal{S}_Z$} & \textover[c]{$0$}{$\mathcal{L}_Z$} & \textover[c]{$0$}{$\mathcal{S}_X^{(M)}$}\\
       \end{matrix}
     \end{array} \right),
   \end{equation}

\noindent where $s$ denotes the syndrome vector; columns two and five correspond to data qubits; and columns three, four, six, and seven correspond to ancilla qubits. After propagating these operators through the circuit, they become

    \begin{equation}
       \left( 
       \begin{array}{@{}c@{}|c@{}|c@{}}
           \begin{matrix}
            \textover[c]{$s$}{$\mathcal{S}_X$}\\\vspace{2pt}
            \textover[c]{$0$}{$\mathcal{Z}$}\\\vspace{2pt}
            \textover[c]{$0$}{${\mathcal{S}}^{M}_X$}\\
            \textover[c]{$0$}{${\mathcal{S}}^{M}_X$}\\
            \textover[c]{$0$}{$\mathcal{S}_X$}\\
            \textover[c]{$0$}{$\mathcal{L}_X$}
           \end{matrix}
       &
            \begin{matrix}
                \mathcal{S}_X & 0 & 0\\
               0&0&0\\
               {\mathcal{S}}_X^{(LM)} & {\mathcal{S}}_X^{(LM)} & {\mathcal{S}}_Z^{(LM)} \\
              \mathcal{S}_X^{(M)} & \mathcal{S}_X^{(M)} & \mathcal{S}_Z^{(M)} \\
              0 & 0 & 0 \\
              0 & 0 & 0
            \end{matrix}
       &
           \begin{matrix}
              \mathcal{S}_Z & \mathcal{S}_Z & \mathcal{S}_X\\
              0 & \multicolumn{2}{c}{\mathcal{Z}^{(M)}}\\
              {\mathcal{S}}_Z^{(LM)} & 0 & 0 \\
              \mathcal{S}_Z^{(M)} & 0 & 0 \\
              0 & \mathcal{S}_Z & \mathcal{S}_X \\
              0 & \mathcal{L}_Z & \mathcal{L}_X
           \end{matrix}
       \\\hline
                \begin{matrix}
                   \phantom{0}
                \end{matrix}
          &
                  \begin{matrix}
                  \textover[c]{$\mathcal{L}_X$}{$\mathcal{S}_X^{(LM)}$} & \textover[c]{$0$}{$\mathcal{S}_X^{(LM)}$} & \textover[c]{$0$}{$\mathcal{S}_Z^{(LM)}$}\\
               \end{matrix}
       &
           \begin{matrix}
             \textover[c]{$\mathcal{L}_Z$}{$\mathcal{S}_Z^{(LM)}$} & \mathcal{L}_Z & \mathcal{L}_X\\
           \end{matrix}
           \end{array} \right)
    \cong
    \left( \begin{array}{@{}c@{}|c@{}|c@{}}
        \begin{matrix}
            s\\\vspace{2pt}
            0\\\vspace{2pt}
            {s}^{(LM)}\\
            s^{(M)}\\
            0\\
            0
        \end{matrix}
       &
       \begin{matrix}
           \mathcal{S}_X & 0 & 0\\
           0&0&0\\\vspace{2pt}
           0 & {\mathcal{S}}_X^{(LM)} & {\mathcal{S}}_Z^{(LM)}\\
          0 & \mathcal{S}_X^{(M)} & \mathcal{S}_Z^{(M)}\\
          0 & 0 & 0 \\
          0 & 0 & 0
       \end{matrix}
       &
       \begin{matrix}
          \mathcal{S}_Z & 0 & 0\\
          0 & \multicolumn{2}{c}{\mathcal{Z}^{(M)}}\\
          0 & 0 & 0 \\
          0 & 0 & 0 \\
          0 & \mathcal{S}_Z & \mathcal{S}_X \\
          0 & \mathcal{L}_Z & \mathcal{L}_X
       \end{matrix}\\
                  \hline
                  \begin{matrix}
                      \phantom{0}
                  \end{matrix}
                  &
                       \begin{matrix}
           \textover[c]{$\mathcal{L}_X$}{$\mathcal{S}_X$} & \textover[c]{$0$}{$\mathcal{S}_X^{(LM)}$} & \textover[c]{$0$}{$\mathcal{S}_Z^{(LM)}$}\\
       \end{matrix}
       &
       \begin{matrix}
           \textover[c]{$\mathcal{L}_Z$}{$\mathcal{S}_X$} & \textover[c]{$0$}{$\mathcal{S}_X$} & \textover[c]{$0$}{$\mathcal{S}_Z$}\\
       \end{matrix}
     \end{array} \right)
\end{equation}

\noindent where $s^{(M)}$ and ${s}^{(LM)}$ denote the syndrome bits associated with the stabilizers $\mathcal{S}^{(M)}$ and ${\mathcal{S}}^{(LM)}$ respectively. Thus, observe that measuring the ancilla in $X$ basis reveals information only on $s^{(M)}$ and ${s}^{(LM)}$---the syndrome bits of $\mathcal{S}^{(M)} \cup {\mathcal{S}}^{(LM)}$, and that the circuit leaves the data unchanged, as intended.

    Moreover, notice in the circuit that the ancilla qubits indexed by $\mathcal{C}^{(M)}$ act trivially throughout due to the stabilizers $\mathcal{Z}^{(M)}$, which consist of single-qubit Pauli-$Z$ operators supported on each such qubit. Consequently, those qubits and their gates can be removed from the circuit without consequence.

\end{proof}

\newpage

\section{On recovering Shor syndrome extraction} \label{shorsyndextrresult}

\begin{theorem}
Let $\mathcal{S}$ be a set of stabilizer generators for some code. The ancilla stabilizer state $\ket{a}$ of our procedure is given in Eq.\ \ref{ExplanationAncillaStabs}, where the stabilizers are partitioned such that $\mathcal{S}=\mathcal{S}^{(M)}\cup \mathcal{S}^{(L)}$ and $\langle \mathcal{S}^{(L)}\rangle = \langle \mathcal{S}^{(LM)}\cup \mathcal{S}^{(LL)} \rangle$ where $\mathcal{S}^{(M)}$ is chosen arbitrarily; ${\mathcal{S}}^{(LM)}$ has zero columns in at least the same locations as $\mathcal{S}^{(M)}$; every row of ${\mathcal{S}}^{(LL)}$ has an entry of 1 in a column that no other row of $\mathcal{S}^{(M)}\cup{\mathcal{S}}^{(LL)}$ has a 1 in; and the $X$ and $Z$ subscripts are defined in Eq.\ \ref{stabilzierdecomp}. (Also, let $\mathcal{C}^{(M)}$ denote the indices of zero columns in $\mathcal{S}^{(M)}$ and $\mathcal{Z}^{(M)}\equiv\{Z_\sigma | \sigma \in {\mathcal{C}^{(M)}}\}$.)

When $\lvert{{\mathcal{S}}^{(LM)}}\rvert=0$ and $\lvert{\mathcal{S}^{(M)}}\rvert=1$, where $\mathcal{S}^{(M)}$ contains a weight-$w$ stabilizer, the ancilla consists of $w$ qubits and is stabilized by $X_1X_2 \dots X_w$. (If the stabilizer $S^{(M)}\in\mathcal{S}^{(M)}$ supports Pauli-$Y$ operators, the version of our procedure employing controlled-$Y$ gates is required.)    
\end{theorem}

There is an intuitive way to see why this result holds. Recall the role of the Pauli-$Z$ operators in the ancilla's stabilizer that also contains Pauli-$X$s: after the Pauli-$X$s supported in the ancilla propagate to the data through $\text{C}[X]$ gates, the subsequent $\text{C}[Z]$ gates propagate Pauli-$Z$s back to the ancilla; the Pauli-$Z$ component of the ancilla's stabilizer is designed so that these Pauli-$Z$s cancel to identity after propagation, leaving the ancilla with some $X$-type stabilizers by the end of the circuit, whose $\pm1$ phase can be read by measuring in the $X$-basis. In the special case of $\lvert{\mathcal{S}^{(M)}}\rvert=1$ and $\lvert{{\mathcal{S}}^{(LM)}}\rvert=0$, the circuit has depth 1. This means no Pauli-$Z$s propagate back to the ancilla, so there is no Pauli-$Z$ component to the ancilla stabilizers containing Pauli-$X$s.

We now give a rigorous proof. Note, as above, the argument hinges on the circuit having depth-1. To ensure this, the proof invokes a detail of the controlled-$Y$-gate variant of our procedure, which has not yet been described. In particular, for each qubit $u$ on which the measured weight-$w$ stabilizer supports a Pauli-$Y$:

\begin{itemize}
    \item Replace the $X_uY_{w+u}$ components of the ancilla stabilizers with $X_u$.
    \item Delete qubit $w+u$.
    \item Replace the controlled-$X$ gate on qubit $u$ with a controlled-$Y$ gate.
\end{itemize}

\noindent (These effectively correspond to the relevant steps in Appendix \ref{CYgateresults}, recalling $\mathcal{Z}^{(M)}_S=0$.)
  
\begin{proof} Let ${S}^{(M)}\in \mathcal{S}^{(M)}$ be the (single) stabilizer in $\mathcal{S}^{(M)}$. By construction, the ancilla is stabilized by $({S}^{(M)}_X \quad {S}^{(M)}_Z \quad|\quad 0\quad {S}^{(M)}_X)$. 

Per the procedure, all qubits corresponding to zero columns of $({S}^{(M)}_X \quad {S}^{(M)}_Z)$ are deleted. After this deletion, the ancilla stabilizer $({S}^{(M)}_X \quad {S}^{(M)}_Z \quad |\quad 0  \quad {S}^{(M)}_X)$ has a left half consisting entirely of 1s. Equivalently, the ancilla is stabilized by an operator that is a tensor product of Pauli-$X$s and Pauli-$Y$s on every remaining qubit.

To identify the qubit sites where the Pauli-$Y$s act in the ancilla stabilizer, examine how deleting qubits with zero columns in ${S}^{(M)}_Z$ affects the right half of the ancilla qubits' stabilizer: $( {S}^{(M)}_Z     \quad|\quad         {S}^{(M)}_X )$. Notice that Pauli-$Y$ support appears exactly on the qubits that neither ${S}^{(M)}_Z$ nor ${S}^{(M)}_X$ had a zero column, i.e., precisely the qubits where $S^{(M)}$ supported a Pauli-$Y$. 

Thus, if $S^{(M)}$ contains $v\leq w$ Pauli-$Y$ operators, then the ancilla is a $(w+v)$-qubit state stabilized by 

\begin{equation}
\left( \begin{array}{@{}c@{}|c@{}}       \begin{matrix}           {S}^{(M)}_X & {S}^{(M)}_Z       \end{matrix}       &       \begin{matrix}          0  & {S}^{(M)}_X      \end{matrix}     \end{array} \right) \cong  X_1X_2\dots X_w Y_{w+1} Y_{w+2} \dots Y_{w+v},
\end{equation}

\noindent up to relabeling of qubits. In the controlled-$Y$ version of our procedure, we map $X_uY_{w+u}\rightarrow X_u$ and delete qubit $w+u$ for all $u\in\{1,2,\dots,v\}$. After this step, the stabilizer $\left( \begin{array}{@{}c@{}|c@{}}       \begin{matrix}           {S}^{(M)}_X & {S}^{(M)}_Z       \end{matrix}       &       \begin{matrix}          0  & {S}^{(M)}_X      \end{matrix}     \end{array} \right) \rightarrow X^{\otimes w}$ and only $w$ qubits remain in the system, completing the proof.

\end{proof}

\newpage

\section{Cat states in Shor syndrome extraction} \label{LemmaFormForStabs}

\begin{theorem}\label{catstatetheorem}
    Let $w\geq2$. Fix a $w$-qubit state to have a stabilizer generator $X_1 X_2 \dots X_w$ and $w-1$ other stabilizer generators that are $Z$-type and have a $+1$ phase. Then the state must have the stabilizers

    \begin{equation}
        \langle X_1 X_2 \dots X_w, Z_1 Z_2, Z_2Z_3, \dots, Z_{w-1}Z_w\rangle.
    \end{equation}

\end{theorem}

\begin{proof}

    For the $Z$-type generators to commute with the stabilizer $X_1 X_2 \dots X_w$, they must have even weight. There are $2^w$ possible $Z$-type, $+1$ phase stabilizers on $w$ qubits. Half, i.e., $2^{w-1}$ total, have even weight. 
  
   Observe that our $w-1$ independent $Z$-type generators generate $2^{w-1}$ operators per Lemma \ref{catstatelemma1}. Therefore, our generators necessarily saturate the entire space of permissible operators. That is, there is only one group our $w-1$ $Z$-type stabilizers can generate: those with $+1$ phase and even weight.

   Concretely, one choice of generators for the even-parity, +1 phase, $Z$-type stabilizer group is 
   
   \begin{equation}
       \langle Z_1 Z_2, Z_2Z_3, \dots, Z_{w-1}Z_w\rangle
   \end{equation}

   \noindent as shown in Lemma \ref{catstatelemma2}.

\end{proof}

\begin{lemma}\label{catstatelemma1}
    Any set of $m$ independent $Z$-type operators $g_1,g_2,\dots, g_m$ generates a group of order $2^m$. 
   \end{lemma}
   
   \begin{proof}

   While this result follows from the definition of independence over the field $\mathbb{Z}_2$, we will also provide a direct proof.

   The set $\langle g_1,g_2,\dots, g_m\rangle$ forms a group under multiplication: there exists an identity $g_1^2$; every element has an inverse, because all elements are self-inverse; associativity is inherited from multiplication; and closure is trivial since we define the group such that it contains all operators $\langle g_1,g_2,\dots, g_m \rangle$.
   
   To compute the group's order, we will follow an argument similar to one in \cite{Lidar_Brun_2013}.  
   Observe, by definition, that any element of the group can be written as a product $g^{a_1}_1 g^{a_2}_2 \dots g^{a_{m}}_{m}$ for some binary vector $a \in \mathbb{Z}_2^{m}$. Suppose two such products yield the same operator:

   \begin{equation}
   g^{a_1}_1 g^{a_2}_2 \dots g^{a_{m}}_{m}=g^{b_1}_1 g^{b_2}_2 \dots g^{b_{m}}_{m}
   \label{linearcombine}
   \end{equation}

   \noindent for some binary vectors $a,b \in \mathbb{Z}_2^{m}$. From commutativity of the generators, this implies that

   \begin{equation}
       g^{a_1\oplus b_1}_1 g^{a_2\oplus b_2}_2 \dots g^{a_{m}\oplus b_{m}}_{m}=I 
   \end{equation}
   
   \noindent and so $a=b$. Hence, each $a \in \mathbb{Z}_2^{m}$ defines a unique group element  $g^{a_1}_1 g^{a_2}_2 \dots g^{a_{m}}_{m}$, and since there are $2^m$ such vectors $a$, the group has order $2^m$.
   \end{proof}

   \begin{lemma}\label{catstatelemma2}
    A generating set for the group of all even-parity, $Z$-type operators with $+1$ phase, on $w$ qubits is

    \begin{equation}
       \{ Z_1 Z_2, Z_2Z_3, \dots, Z_{w-1}Z_w\}.
   \end{equation}
   \end{lemma}
   
    \begin{proof} From Lemma \ref{catstatelemma1}, we know that if the proposed generators are independent, then they generate a group of size $2^{w-1}$. 
    Additionally, as shown in Theorem \ref{catstatetheorem}, there are exactly $2^{w-1}$ even-parity $Z$-type operators with a $+1$ phase on $w$ qubits. Clearly, the stabilizers generated by the proposed operators will all have a $+1$ phase. 
    Therefore, if we can show that the proposed operators are all independent and only generate even-weight operators, the claim follows, since then they saturate the space of even-weight $Z$-type operators with a $+1$ phase.

    To establish independence, see that the proposed generators can be rewritten, up to products of one another, as 

        \begin{equation}
       \{ Z_1 Z_2, Z_1Z_3, \dots, Z_1Z_w\}.
   \end{equation}

 \noindent In this form, each operator has support on a unique qubit, so they cannot be expressed as a product of the others, and are thus independent.

     Next, notice that the product of any two even-weight, $Z$-type operators is again an even-weight, $Z$-type operator. This follows from breaking the multiplication of two such operators into two steps: first, consider qubits where their supports intersect, and second, those qubits where their supports are disjoint. In particular,

     \begin{itemize}
         \item On qubits where both operators support $Z$s, the $Z$s cancel pairwise (because $Z^2=I$), leaving two $Z$-type operators with weights that sum to an even number and where their supports are now on a disjoint set of qubits.
         \item The product of two $Z$-type operators that have support on a disjoint set of qubits and where the sum of their weights is even will support all the $Z$s from both operators, of which there is an even number
     \end{itemize}

    \noindent The proof is completed by seeing that all the generators have even weight, so any product of them too must have an even weight.

    \end{proof}

\newpage
\section{Using controlled-\texorpdfstring{$Y$}{Y} gates}
\label{CYgateresults}

We now describe how our standard procedure is adapted to use controlled-$Y$ gates. Roughly speaking, if the $n$-qubit stabilizer generators being measured, ${\mathcal{S}}^{(M)}\cup {\mathcal{S}}^{(LM)}$, act only with Pauli-$Y$s or identities on certain qubits, the ancilla is changed such that the corresponding $\text{C}[X]$ and $\text{C}[Z]$ gates (controlled on different ancilla qubits) are replaced by $\text{C}[Y]$ gates (each controlled on a single ancilla qubit). 

Formally, we use the circuit in Figure \ref{fig:ShorAndSteane}c, where each qubit of the third ancilla register is tentatively prepared in a $+1$ eigenstate of Pauli-$Z$. That is, the ancilla stabilizers of our standard procedure, Eq.\ \ref{AncillaStabsProcedure}, are replaced by

    \begin{equation}
       \left( \begin{array}{@{}c@{}|c@{}}
       \begin{matrix}
          \mathcal{S}_X & \mathcal{S}_Z & 0 \\
          0 & 0 & 0\\
          0 & 0 & 0\\
          0 & 0 & 0\\

       \end{matrix}
       &
       \begin{matrix}
          0 & \mathcal{S}_X& 0 \\
          \mathcal{S}_Z & \mathcal{S}_X& 0 \\
          \mathcal{L}_Z & \mathcal{L}_X& 0\\
          0&0&\mathcal{Z}_Y
       \end{matrix}
       \end{array} \right)
=     
       \left( \begin{array}{@{}c@{}|c@{}}
       \begin{matrix}
          {\mathcal{S}}_X^{(LL)} & {\mathcal{S}}_Z^{(LL)}&0 \\
          {\mathcal{S}}_X^{(LM)} & {\mathcal{S}}_Z^{(LM)}&0 \\
          \mathcal{S}_X^{(M)} & \mathcal{S}_Z^{(M)}&0 \\
          0 & 0 &0\\
          0 & 0&0 \\
          0&0&0
       \end{matrix}
       &
       \begin{matrix}
          0 & {\mathcal{S}}_X^{(LL)}&0 \\
          0 & {\mathcal{S}}_X^{(LM)} &0\\
          0 & \mathcal{S}_X^{(M)} &0\\
          \mathcal{S}_Z & \mathcal{S}_X&0 \\
          \mathcal{L}_Z & \mathcal{L}_X &0\\
          0&0&\mathcal{Z}_Y
       \end{matrix}
       \end{array} \right).
   \end{equation}

\noindent where $\mathcal{Z}_Y$ denotes a single-qubit Pauli-$Z$ on each qubit of the third block. At this stage, all the $\text{C}[Y]$ gates in the circuit of Figure \ref{fig:ShorAndSteane}c are trivial. The standard procedure is then applied, giving an ancilla stabilized by

\begin{equation}
    \left( \begin{array}{@{}c@{}|c@{}}
       \begin{matrix}
       0&0&0\\
          {\mathcal{S}}_X^{(LM)} & {\mathcal{S}}_Z^{(LM)}&0 \\
          \mathcal{S}_X^{(M)} & \mathcal{S}_Z^{(M)}&0 \\
          0 & 0 &0\\
          0 & 0&0 \\
          0&0&0
       \end{matrix}
       &
       \begin{matrix}
       \multicolumn{2}{c}{\mathcal{Z}^{(M)}} & 0\\
          0 & {\mathcal{S}}_X^{(LM)} &0\\
          0 & \mathcal{S}_X^{(M)} &0\\
          \mathcal{S}_Z & \mathcal{S}_X&0 \\
          \mathcal{L}_Z & \mathcal{L}_X &0\\
          0&0&\mathcal{Z}_Y
       \end{matrix}
       \end{array} \right),
       \label{CYmatrixforstabs}
\end{equation}

\noindent recalling that $\mathcal{Z}^{(M)}$ contains rows with a single 1 in each of the zero columns of $\mathcal{S}^{(M)}$. 

The controlled-$Y$ modification consists of the following adjustments. Let set $\mathcal{Y}\subseteq\{1,2,\dots,n\}$ contain the qubits on which every generator in ${\mathcal{S}}^{(M)}\cup {\mathcal{S}}^{(LM)}$ acts only with Pauli-$Y$s or identities,  with at least one Pauli-$Y$ on each such qubit. Then, for all $j\in\mathcal{Y}$:

    \begin{enumerate}
        \item Replace the row corresponding to $Z_{2n+j}$ (in the final block of rows) with $Z_j$. (As shown below, the resulting stabilizer group also contains $Z_{n+j}$.)

        \item   For each row in the ancilla's second and third blocks of rows:
            \begin{enumerate}
                \item If the corresponding operator supports $X_j$ xor $X_{n+j}$, multiply the row operator by $Z_{2n+j}$. In practice, this step is trivial: every stabilizer in ${\mathcal{S}}^{(M)}\cup {\mathcal{S}}^{(LM)}$ supports either a $Y$ or an identity on qubits $j\in\mathcal{Y}$. Therefore, blocks 2 and 3 of the ancilla stabilizers support either $X_jY_{n+j}$ or act trivially on qubits $j$ and $n+j$. We retain this step nonetheless, since it simplifies the proof.
                \item Else, if that row operator supports $X_j Y_{n+j}$, replace the $X_j Y_{n+j}$ support with $X_{2n+j}$.
            \end{enumerate}

        \item Similarly, for each row operator in the ancilla's fourth and fifth blocks that supports $Z_j$ xor $Z_{n+j}$, multiply that stabilizer by $Z_{2n+j}$.

    \end{enumerate}

    \noindent After the loop over $\mathcal{Y}$ is completed, the ancilla stabilizers become

\begin{equation}
       \left( \begin{array}{c@{}|c@{}}

       \begin{matrix}
          0&0&0 \\
          {\Tilde{\mathcal{S}}}_X^{(LM)} & {\Tilde{\mathcal{S}}}_Z^{(LM)}&\mathcal{X}^{(LM)} \\
          \Tilde{\mathcal{S}}_X^{(M)} & \Tilde{\mathcal{S}}_Z^{(M)}&\mathcal{X}^{(M)} \\
          0 & 0 &0\\
          0 & 0&0 \\
          0&0&0\\
          0&0&0\\
          0&0&0
       \end{matrix}
       &
       \begin{matrix}
           \multicolumn{2}{c}{\mathcal{Z}^{(M)}}&0 \\
          0 & \Tilde{{\mathcal{S}}}_X^{(LM)} &\mathcal{Z}^{(LM)}_S\\
          0 & \Tilde{\mathcal{S}}_X^{(M)} &\mathcal{Z}^{(M)}_S\\
          \mathcal{S}_Z & \mathcal{S}_X& \mathcal{Z}_S \\
          \mathcal{L}_Z & \mathcal{L}_X & \mathcal{Z}_L\\
          {\mathcal{Z}}_X&0&0\\
          0&{\mathcal{Z}}_Z&0\\
          0&0&\Tilde{\mathcal{Z}}_Y\\
       \end{matrix}
       \end{array} \right)
       \label{nextCYmatrixforstabs}
   \end{equation}

\noindent where the tildes on ${{\mathcal{S}}}_X^{(LM)}$, ${{\mathcal{S}}}_Z^{(LM)}$, ${{\mathcal{S}}}_X^{(M)}$, and ${{\mathcal{S}}}_Z^{(M)}$, together with the terms $\mathcal{X}^{(LM)}$ and $\mathcal{X}^{(M)}$, encode the modification in step 2 where $X_j Y_{n+j}\rightarrow X_{2n+j}$; the matrices $\mathcal{Z}_S$ and $\mathcal{Z}_L$ capture the changes in step 3, while $\mathcal{Z}^{(LM)}_S=0$ and $\mathcal{Z}^{(M)}_S=0$ hold the rows of $\mathcal{Z}$ corresponding to the stabilizers $\mathcal{S}^{(LM)}$ and $\mathcal{S}^{(M)}$, respectively; the sixth and seventh blocks respectively represent the operators $\{Z_j : j\in\mathcal{Y}\}$ and $\{Z_{n+j} : j\in\mathcal{Y}\}$, introduced in step 1; and the final block of rows corresponds to $\{Z_{2n+j} : j\in\mathbb{Z}_n\land j\notin\mathcal{Y}\}$, also from step 1. Conceptually, the tildes mark \textit{holes} in the former operators---on qubits supporting only Pauli-$Y$s or identities---while some of the new terms can be viewed as filling these holes on the additional register.

Intuitively, steps 1 and 2 render the $\text{C}[Y]$ gates on qubit $j$ nontrivial, while simultaneously making the corresponding $\text{C}[X]$ and $\text{C}[Z]$ gates trivial. The $Z_{n+j}$ factors (included in the $Y_{n+j}$) in step 2 are removed because the associated $\text{C}[Z]$ gates are deleted, so cancellation of unwanted propagation on that qubit is no longer needed. The $Z_{2n+j}$ factors introduced in steps 2 and 3 anticipate how the data-ancilla system's stabilizers and logical operators propagate through the nontrivial $\text{C}[Y]$ gates, ensuring that unwanted propagation is canceled.

Step 4: delete the following qubits, along with any gates acting on them: qubits $j$ and $n+j$ for all $j\in\mathcal{Y}$; qubits $2n+k$ for all $k\notin \mathcal{Y}$; qubits $l\in\mathcal{C}^{(M)}\subseteq\{1,2,\dots,2n\}$, recalling $\mathcal{C}^{(M)}$ denotes the indices of zero columns in $\mathcal{S}^{(M)}=(\mathcal{S}^{(M)}_X\mid\mathcal{S}^{(M)}_Z)$. Equivalently, delete all qubits with nontrivial support in the rows containing ${\mathcal{Z}}_X$, ${\mathcal{Z}}_Z$, $\Tilde{\mathcal{Z}}_Y$, and $\mathcal{Z}^{(M)}$.

After applying these steps, the ancilla remains a valid stabilizer state and still extracts the syndrome bits of ${\mathcal{S}}^{(M)}\cup {\mathcal{S}}^{(LM)}$, as shown in the following two theorems.

\begin{theorem}
The ancilla and circuit extract only the syndrome bits of the generators ${\mathcal{S}}^{(M)}\cup {\mathcal{S}}^{(LM)}$. 
    
\end{theorem}

\noindent The analogous result for our standard procedure is given in Appendix \ref{ProofSyndromeExtraction}.

\begin{proof} 

Recall that, before deleting qubits, the stabilizers of the ancilla are given in Eq.\ \ref{nextCYmatrixforstabs}. That is, the stabilizers and logical operators of the data-ancilla system are

    \begin{equation}
       \left( \begin{array}{@{}c@{}|c@{}|c@{}}
       \begin{matrix}
           s\\0\\0\\0\\0\\0\\0\\0\\0
       \end{matrix}
       &
       \begin{matrix}
       \mathcal{S}_X&0&0&0\\
          0&0&0&0 \\
          0&{\Tilde{\mathcal{S}}}_X^{(LM)} & {\Tilde{\mathcal{S}}}_Z^{(LM)}&\mathcal{X}^{(LM)} \\
          0&\Tilde{\mathcal{S}}_X^{(M)} & \Tilde{\mathcal{S}}_Z^{(M)}&\mathcal{X}^{(M)} \\
          0&0 & 0 &0\\
          0&0 & 0&0 \\
          0&0&0&0\\
          0&0&0&0\\
          0&0&0&0
       \end{matrix}
       &
       \begin{matrix}
       \mathcal{S}_Z&0&0&0\\
          0&\multicolumn{2}{c}{\mathcal{Z}^{(M)}}&0 \\
          0&0 & \Tilde{{\mathcal{S}}}_X^{(LM)} &\mathcal{Z}^{(LM)}_S\\
          0&0 & \Tilde{\mathcal{S}}_X^{(M)} &\mathcal{Z}^{(M)}_S\\
          0&\mathcal{S}_Z & \mathcal{S}_X& \mathcal{Z}_S \\
          0&\mathcal{L}_Z & \mathcal{L}_X & \mathcal{Z}_L\\
          0&{\mathcal{Z}}_X&0&0\\
          0&0&{\mathcal{Z}}_Z&0\\
          0&0&0&\Tilde{\mathcal{Z}}_Y\\
       \end{matrix}\\
       \hline
       \begin{matrix}
          \phantom{0}
           \end{matrix}
           &
           \begin{matrix}
                          \textover[c]{$\mathcal{L}_X$}{$\mathcal{S}_X$} & 
                          \textover[c]{$0$}{$\tilde{\mathcal{S}}_X^{(LM)}$} & 
                          \textover[c]{$0$}{$\tilde{\mathcal{S}}_Z^{(LM)}$}&
                          \textover[c]{$0$}{$\mathcal{X}^{(LM)}$}\\
       \end{matrix}
       &
       \begin{matrix}
           \textover[c]{$\mathcal{L}_Z$}{$\mathcal{S}_Z$} & 
           \textover[c]{$0$}{$\mathcal{Z}_X$} & 
           \textover[c]{$0$}{$\Tilde{\mathcal{S}}_X^{(LM)}$}&
           \textover[c]{$0$}{$\mathcal{Z}_S^{(LM)}$}\\
       \end{matrix}
       \end{array} \right)
   \end{equation}

\noindent where $s$ denotes the syndrome vector; columns two and six correspond to data qubits; and the remaining columns correspond to ancilla qubits.

 Propagating these operators through the circuit in Figure \ref{fig:ShorAndSteane}c, they become

\begin{equation}
       \left( \begin{array}{@{}c@{}|c@{}|c@{}}
       \begin{matrix}
           s\\0\\0\\0\\0\\0\\0\\0\\0
       \end{matrix}
       &
       \begin{matrix}
       \mathcal{S}_X&0&0&0\\
          0&0&0&0 \\
          {{\mathcal{S}}}_X^{(LM)}&{\Tilde{\mathcal{S}}}_X^{(LM)} & {\Tilde{\mathcal{S}}}_Z^{(LM)}&\mathcal{X}^{(LM)} \\
          {\mathcal{S}}_X^{(M)}&\Tilde{\mathcal{S}}_X^{(M)} & \Tilde{\mathcal{S}}_Z^{(M)}&\mathcal{X}^{(M)} \\
          0&0 & 0 &0\\
          0&0 & 0&0 \\
          0&0&0&0\\
          0&0&0&0\\
          0&0&0&0
       \end{matrix}
       &
       \begin{matrix}
       \mathcal{S}_Z&\mathcal{S}_Z&\mathcal{S}_X&\mathcal{S}_X\oplus \mathcal{S}_Z\\
          0& \multicolumn{2}{c}{\mathcal{Z}^{(M)}}&0 \\
          {{\mathcal{S}}}_Z^{(LM)}&0 & 0&\Tilde{{\mathcal{S}}}_X^{(LM)}\oplus\Tilde{{\mathcal{S}}}_Z^{(LM)}\oplus\mathcal{Z}^{(LM)}_S\\
          {\mathcal{S}}_Z^{(M)}&0 & 0 &\Tilde{\mathcal{S}}_X^{(M)} \oplus \Tilde{\mathcal{S}}_Z^{(M)}\oplus\mathcal{Z}^{(M)}_S\\
          0&\mathcal{S}_Z & \mathcal{S}_X& \mathcal{Z}_S \\
          0&\mathcal{L}_Z & \mathcal{L}_X & \mathcal{Z}_L\\
          0&{\mathcal{Z}}_X&0&0\\
          0&0&{\mathcal{Z}}_Z&0\\
          0&0&0&\Tilde{\mathcal{Z}}_Y\\
       \end{matrix}\\
       \hline
       \begin{matrix}
          \phantom{0}
           \end{matrix}
           &
           \begin{matrix}
                          \textover[c]{$\mathcal{L}_X$}{${\Tilde{\mathcal{S}}}_X^{(LM)} $} & \textover[c]{$0$}{${\Tilde{\mathcal{S}}}_X^{(LM)} $} & \textover[c]{$0$}{${\Tilde{\mathcal{S}}}_X^{(LM)} $}&\textover[c]{$0$}{$\mathcal{X}^{(LM)}$}\\
       \end{matrix}
       &
       \begin{matrix}
           \textover[c]{$\mathcal{L}_Z$}{${\Tilde{\mathcal{S}}}_X^{(LM)}$} & \textover[c]{$\mathcal{L}_Z$}{$\mathcal{L}_Z$} & \textover[c]{$\mathcal{L}_X$}{$\mathcal{L}_Z$}&\textover[c]{$\mathcal{L}_X \oplus \mathcal{L}_Z$}{$\Tilde{{\mathcal{S}}}_X^{(LM)}\oplus\Tilde{{\mathcal{S}}}_Z^{(LM)}\oplus\mathcal{Z}^{(LM)}_S$}\\
       \end{matrix}
       \end{array} \right)
   \end{equation}

\noindent having used the relations

\begin{equation}
    {\Tilde{{\mathcal{S}}}}_X^{(LM)}\oplus\mathcal{X}^{(LM)}={{\mathcal{S}}}_X^{(LM)}, \quad {\Tilde{{\mathcal{S}}}}_Z^{(LM)}\oplus\mathcal{X}^{(LM)}={{\mathcal{S}}}_Z^{(LM)}, 
 \end{equation}   
  \begin{equation}  
    {\Tilde{{\mathcal{S}}}}_X^{(M)}\oplus\mathcal{X}^{(M)}={{\mathcal{S}}}_X^{(M)}, \quad{\Tilde{{\mathcal{S}}}}_Z^{(M)}\oplus\mathcal{X}^{(M)}={{\mathcal{S}}}_Z^{(M)},
\end{equation}

\noindent where, for completeness, $\oplus$ denotes matrix addition modulo 2. Up to products of stabilizers, they are equivalent to

\begin{equation}
       \left( \begin{array}{@{}c@{}|c@{}|c@{}}
       \begin{matrix}
           s\\0\\{s}^{(LM)}\\{s}^{(M)}\\0\\0\\0\\0\\0
       \end{matrix}
       &
       \begin{matrix}
       \mathcal{S}_X&0&0&0\\
          0&0&0&0 \\
          0&{\Tilde{\mathcal{S}}}_X^{(LM)} & {\Tilde{\mathcal{S}}}_Z^{(LM)}&\mathcal{X}^{(LM)} \\
          0&\Tilde{\mathcal{S}}_X^{(M)} & \Tilde{\mathcal{S}}_Z^{(M)}&\mathcal{X}^{(M)} \\
          0&0 & 0 &0\\
          0&0 & 0&0 \\
          0&0&0&0\\
          0&0&0&0\\
          0&0&0&0
       \end{matrix}
       &
       \begin{matrix}
       \mathcal{S}_Z&0&0&0\\
          0& \multicolumn{2}{c}{\mathcal{Z}^{(M)}}&0 \\
          0&0 & 0&0\\
          0&0 & 0 &0\\
          0&\mathcal{S}_Z & \mathcal{S}_X& \mathcal{Z}_S \\
          0&\mathcal{L}_Z & \mathcal{L}_X & \mathcal{Z}_L\\
          0&{\mathcal{Z}}_X&0&0\\
          0&0&{\mathcal{Z}}_Z&0\\
          0&0&0&\Tilde{\mathcal{Z}}_Y\\
       \end{matrix}\\
       \hline
       \begin{matrix}
          \phantom{0}
           \end{matrix}
           &
           \begin{matrix}
                          \textover[c]{$\mathcal{L}_X$}{${\Tilde{\mathcal{S}}}_X$} & \textover[c]{$0$}{${\Tilde{\mathcal{S}}}_X^{(LM)} $} & \textover[c]{$0$}{${\Tilde{\mathcal{S}}}_X^{(LM)} $}&\textover[c]{$0$}{$\mathcal{X}^{(LM)}$}\\
       \end{matrix}
       &
       \begin{matrix}
           \textover[c]{$\mathcal{L}_Z$}{${\Tilde{\mathcal{S}}}_Z$} &
           \textover[c]{$0$}{$\mathcal{L}_Z$} &
           \textover[c]{$0$}{$\mathcal{L}_Z$}&
           \textover[c]{$0$}{$\mathcal{Z}_Y$}\\
       \end{matrix}
       \end{array} \right)
   \end{equation}

\noindent where $s^{(LM)}$ and ${s}^{(M)}$ denote the syndrome bits of the stabilizers $\mathcal{S}^{(LM)}$ and ${\mathcal{S}}^{(M)}$, respectively, and this is justified by the following observations:

\begin{equation}
    \Tilde{\mathcal{S}}_X^{(LM)} \oplus \Tilde{\mathcal{S}}_Z^{(LM)}={\mathcal{S}}_X^{(LM)} \oplus {\mathcal{S}}_Z^{(LM)}, \quad
\Tilde{\mathcal{S}}_X^{(M)} \oplus \Tilde{\mathcal{S}}_Z^{(M)}={\mathcal{S}}_X^{(M)} \oplus {\mathcal{S}}_Z^{(M)}, 
\end{equation}
\begin{equation}
    \mathcal{S}_X \oplus \mathcal{S}_Z \oplus \mathcal{Z}_S\subseteq\langle \Tilde{\mathcal{Z}}_Y \rangle,\quad
     \mathcal{L}_X \oplus \mathcal{L}_Z \oplus \mathcal{Z}_L\subseteq\langle \Tilde{\mathcal{Z}}_Y \rangle,
\end{equation}

\noindent where the rows of $\Tilde{\mathcal{Z}}_Y$ are interpreted as set elements. The inclusions in this last line follow from: each row of $\mathcal{S}_X\oplus \mathcal{S}_Z$ has a 1 in column $k\in\{1,2,\dots,n\}$ iff the underlying stabilizer operator $S=(S_X \mid S_Z)$ supports an $X$ xor a $Z$ on qubit $k$; each row of $\mathcal{Z}_S$ has a 1 in column $k$ iff $k\in\mathcal{Y}$ and said stabilizer $S$ supports an $X$ xor a $Z$ on qubit $k$; and each row of $\Tilde{\mathcal{Z}}_Y$ has a single 1, located in column $k$ for all $k \notin\mathcal{Y}$. The same holds for $\mathcal{L}_X \oplus \mathcal{L}_Z$ and $\mathcal{Z}_L$.

Therefore, as before, measuring the ancilla in $X$ basis yields $s^{(M)}$ and ${s}^{(LM)}$, i.e.\ the syndrome bits of $\mathcal{S}^{(M)} \cup {\mathcal{S}}^{(LM)}$, while leaving the data unchanged. The syndrome bits are thus extracted as intended.

Moreover, notice that the qubits corresponding to zero columns of $\mathcal{S}^{(M)}$ act trivially throughout the circuit due to the stabilizers corresponding to $\mathcal{Z}^{(M)}$. Similarly, from the inclusion of stabilizers corresponding to $\mathcal{Z}_X$ and $\mathcal{Z}_Z$, the controlled-$X$ and controlled-$Z$ gates we intend to replace by controlled-$Y$ gates also act trivially, as do the unused controlled-$Y$ gates, due to the stabilizers corresponding to $\Tilde{\mathcal{Z}}_Y$. Therefore, those qubits and their gates can be removed from the circuit without consequence.

\end{proof}

\begin{theorem}
 The ancilla forms a valid stabilizer state.
\end{theorem} 

\noindent The analogous result for our standard procedure is given in Appendix \ref{validstabs}.

\begin{proof} 

We begin by showing that the initial stabilizers, before the additional steps in our procedure and given in Eq.\ \ref{CYmatrixforstabs}, commute. This holds because: all but the last block of rows are shown to commute in Appendix \ref{validstabs}; the stabilizers in the last block mutually commute as they are $Z$-type; and the last block commutes with all other stabilizers since they act on a disjoint set of qubits.

We now show, case by case, that by the end of step 3 in the procedure and at any point in the loop, i.e.\ for any $j\in\mathcal{Y}$, commutation is preserved.

Recall that the condition in step 2a of the procedure is never satisfied, and so that step may be trivially disregarded.

The stabilizers considered in step 2 still commute with those corresponding to $\mathcal{Z}^{(M)}$ and the remaining elements (after step 1) of $\mathcal{Z}_Y$. Definitionally, $\mathcal{Z}^{(M)}$ and the remaining elements of $\mathcal{Z}_Y$ have no support on qubits $j$, $j+n$, or $j+2n$. Thus, since these operators commuted with the second and third blocks of rows (the rows changed in step 2) before replacing $X_jY_{j+n}$ with $X_{j+2n}$, they must also commute afterwards.

The stabilizers considered in step 2 still commute with those introduced in step 1 ($Z_j$ and, effectively, $Z_{n+j}$), since the stabilizers from step 2 have no Pauli-$X$ (or Pauli-$Y$) support on qubits $j$ or $j+n$. This follows from the fact that $j\in\mathcal{Y}$ are the qubits for which every stabilizer in ${\mathcal{S}}^{(M)}\cup {\mathcal{S}}^{(LM)}$ supports either a $Y$ or an identity. Consequently, blocks 2 and 3 of the ancilla support either $X_jX_{n+j}$ (effectively removed in step 2) or have no support on qubits $j$ and $n+j$.

The stabilizers considered in steps 2 and 3 still commute. First, consider step-3 stabilizers that support either $I_jI_{j+n}$ or $Z_jZ_{n+j}$. Removing $X_jY_{n+j}$ from the step-2 stabilizers does not affect the commutation relations with these step-3 stabilizers: since they commuted before the steps, they must also commute afterwards. The other case is when the step-3 stabilizers support either $Z_j$ or $Z_{j+n}$. This support would anticommute with any $X_{j}Y_{n+j}$ support in step-2 stabilizers, so just deleting $X_{j}Y_{n+j}$ would mean the two resulting operators would anticommute. However, the step-2 operators also support $X_{2n+j}$, while the step-3 operators also support $Z_{2n+j}$, which also anticommute; the operators must commute overall. The same argument applies to operators ${L}$ in step 3.

The stabilizers considered in step 2 all mutually commute. Let $S$ be a stabilizer considered in step 2, and let $S'$ be that same stabilizer but with its support on qubits $j$, $n+j$, and $2n+j$ removed. Then, for $j\in\mathcal{Y}$ and any given stabilizer row in step 2, there are four possible updates:

\begin{enumerate}
    \item If $S$ has no support on qubits $j$ and $n+j$, then $S\equiv S'\rightarrow S\equiv S'$.
    \item If $S$ has no support on qubit $j$ but supports an $X$ on qubit $n+j$, then $S\equiv S'X_{n+j}\rightarrow SZ_{2n+j}\equiv S'X_{n+j}Z_{2n+j}$.
    \item If $S$ has no Pauli-$X$ support on qubit $n+j$ but supports a Pauli-$X$ on qubit $j$, then $S\equiv S'X_jZ_{n+j}\rightarrow SZ_{2n+j}\equiv S'X_jZ_{n+j}Z_{2n+j}$.
    \item If $S$ supports a $Y$ on qubit $n+j$, then $S\equiv S'X_{j}Y_{n+j}\rightarrow S'X_{2n+j}$.
\end{enumerate}

\noindent (Again, recall that cases 2 and 3 never occur and may be neglected.) To verify that the stabilizers considered in step 2 mutually commute, consider all three choices for pairs of cases 1 and 4, letting $S_1$ and $S_2$ be the pair of stabilizers being considered.

\begin{enumerate}
    \item Two instances of case 1. The resulting operators remain unchanged, so they continue to commute.

    \item Two instances of case 4. That is,

    \begin{align}
        S_1\equiv S'_1X_{j}Y_{n+j}\rightarrow& S'_1X_{2n+j},\\
        S_2\equiv S'_2X_{j}Y_{n+j}\rightarrow& S'_2X_{2n+j}.
    \end{align}

    \noindent Therefore,

\begin{align}
0=[S_1,S_2]=[S_1'X_jY_{n+j},S_2'X_jY_{n+j}]=[S_1',S_2']
  =[S_1'X_{2n+j},S_2'X_{2n+j}].
\end{align}

    \item An instance of case 1 and case 4. That is,
    
        \begin{align}
        S_1\equiv S_1' \rightarrow& S_1\equiv S_1',\\
        S_2\equiv S'_2X_{j}Y_{n+j}\rightarrow& S'_2X_{2n+j}.
    \end{align}
    
    \noindent Therefore,

\begin{align}
 0&=[S_1,S_2]= [S_1',S_2'X_jY_{n+j}]=[S_1',S_2']X_jY_{n+j}\\
 \implies0&=[S_1',S_2']\\
  \implies0&=[S_1',S_2']X_{2n+j}=[S_1',S_2'X_{2n+j}].
\end{align}

\end{enumerate}

\noindent Thus, in all cases, the stabilizers considered in step 2 still mutually commute.

Additionally, the stabilizers introduced in steps 1 and 3 commute with each other and with all other stabilizers. They commute with those of step 2, as shown above; and since the stabilizers in step 1, the changes to the stabilizers in step 3, and the remaining stabilizers are $Z$-type, they necessarily commute.

Step 4 of the procedure also preserves commutativity. Since the stabilizers commute before the step, removing the specified qubits does not change commutativity: each generator can simply be erased of its support on the qubits to be removed (except for single-qubit Pauli-$Z$ stabilizers), and these resulting generators still commute.

We now show that the resulting $3n$-qubit ancilla state (so up to step 4 of the procedure) indeed has $3n$ independent stabilizer generators and therefore no nontrivial logical operators.

Among the initial stabilizers in Eq.\ \ref{CYmatrixforstabs} are $3n$ independent generators: the rows in all but the last block contain $2n$ independent generators, as proven in Appendix \ref{validstabs}, while the last block contains $n$ more independent generators, since each has support disjoint from that of every other row in Eq.\ \ref{CYmatrixforstabs}.

It remains to verify that this independence is preserved throughout the procedure. We do so by fixing any set of $3n$ independent generators from the list above and tracking how they change, case by case.

Each operator considered in step 2 remains independent from all other stabilizers. If a stabilizer in step 2 does not support $X_jY_{n+j}$, it is left unchanged and thus retains its independence. The other possible case is when the stabilizer supports $X_jY_{n+j}$. Here, step 2 maps $X_jY_{n+j} \rightarrow X_{2n+j}$, and the resulting stabilizers remain independent since no other stabilizers besides these support $X_{2n+j}$: in particular, if two stabilizers differed only by $X_jX_{n+j}Z_{n+j}$ before the step, they will still differ by $X_{2n+j}$ after the step, recalling that no operators support $X_{2n+j}$ before this step. In fact, the stabilizers considered in step 2 cannot lose their Pauli-$X$ support by linearly combining rows: if they could, there would exist a nontrivial product of code generators $\mathcal{S}$ mapping to the identity---where squaring an operator is taken as trivial---contradicting Eq.\ \ref{linearcombine} and the surrounding text (a non-zero string $a\neq 00\dots 0$ generates the same operator as string $b= 00\dots 0 \neq a$, which is a contradiction).

Step 3 also preserves independence. The stabilizers considered in step 3 were independent before the step, and multiplying them by at most $Z_{2n+j}$ cannot introduce dependence: no other stabilizers (besides those with $X$-support, from which these $Z$-type stabilizers are certainly independent per above) have $Z$ support on qubit $2n+j$.

We now show that the $\lvert{\mathcal{Y}}\rvert$ operators $\{Z_j\mid j\in \mathcal{Y}\}$ (introduced in step 1) are independent from the other stabilizers and therefore replace the $\lvert{\mathcal{Y}}\rvert$ generators removed in step 1.

Recall that the only $X$-supporting generators are those in step 2, and none can lose their Pauli-$X$ support under row combinations. Hence, no combination involving these $X$-supporting stabilizers from step 2 can yield $Z_j$ for any $j\in \mathcal{Y}$. Therefore, each $Z_j$, where $j\in \mathcal{Y}$, is independent from the stabilizer generators supporting Pauli-$X$s.

What remains to consider is whether the ancilla's $Z$-type stabilizers can generate $\{Z_j\mid j\in \mathcal{Y}\}$. We have already shown that all the ancilla's stabilizers commute, and moreover that commutativity is preserved immediately after step 3 at any point in the procedure's loop over $j\in \mathcal{Y}$. Fix $j\in \mathcal{Y}$ to be any point in the loop and consider the (commuting) stabilizers before step 1. By definition, there is at least one stabilizer supporting an $X_j$. Choose any such operator. This operator commutes with the stabilizer generators considered in step 3---since they commuted before the step, and any additional support introduced in step 3 lies on qubits where this operator has no support---as well as all other $Z$-type stabilizer generators currently in the group (which includes all the $Z$-type stabilizer generators present after step 3, besides $Z_j$). Therefore, none of the $Z$-type stabilizer generators (besides $Z_j$) by the end of step 3 can generate an operator that anticommutes with this $X_j$-supporting operator; $Z_j$ must be independent from the $Z$-type stabilizers. The procedure then continues with a new set of commuting operators; the above argument applies for all $j\in\mathcal{Y}$. Thus, the addition of the stabilizers $\{Z_j\mid j\in\mathcal{Y}\}$ contributes $\lvert{\mathcal{Y}}\rvert$ independent operators, as required.

Note, by Lemma \ref{basicresultonstabs}, the operators $\{Z_{n+j}\mid j\in\mathcal{Y}\}$ are also contained in the group generated by these $3n$ stabilizers, because---as shown above---they commute with all of them.

Finally, after step 4, the ancilla state remains a valid stabilizer state. By multiplying generators, each can be varied to erase its support on the qubits to be removed, except for those consisting solely of a single-qubit Pauli-$Z$ on such qubits. Then, one generator is deleted for each qubit removed, and the deleted stabilizers are necessarily generators since no other stabilizer in the group has support on those qubits. The remaining stabilizers continue to commute as they commuted before deleting qubits, and they had no support on the deleted qubits.

\end{proof}

\newpage
\section{Simulations} \label{sims_appendix}

We used the software tools in \cite{Stim, EncodingPackage} to compare our syndrome-extraction procedures with those of Shor and Steane. The methodology of these simulations is described below.

{
Error-correcting codes were sampled from \cite{codetables}, which catalogs code constructions with the best-known code distances for fixed numbers of physical and logical qubits. The sampling was subject to three constraints: the codes selected were non-CSS, for novelty;} they had few stabilizers, to make decoding with a \textit{look-up table} tractable; and they used the fewest physical qubits at a fixed number of logical qubits and odd code distance, to favor codes with strong guarantees on error-correcting capability at minimal physical qubit cost.

The schemes were evaluated under \textit{circuit-level depolarizing noise with measurement errors}. That is, for a physical error rate $p$:

\begin{itemize}
    \item Following each single-qubit gate or idle location, an error was applied on its support with probability $p$, chosen uniformly from $X$, $Y$, and $Z$;
    \item Following each two-qubit gate, an error was applied on its support with probability $p$, chosen uniformly from the fifteen non-identity two-qubit Pauli operators;
    \item Measurement outcomes were independently flipped with probability $p$;
    \item Data and ancilla preparation were fault-free, with independent single-qubit faults applied in the same way as above immediately after preparation was complete. 
\end{itemize}

We used evenly spaced physical error rates $p$ between $0.05\%$ and $1\%$. These error rates reveal differences in logical failure rates among the schemes while providing useful statistical precision at a computationally tractable cost of roughly $10^5$ shots per data point.

At the end of each syndrome extraction shot, the measured syndrome was decoded and the corresponding correction was applied. We then noiselessly measured the residual syndrome and noiselessly applied its corresponding correction. The reported logical failure rate was the fraction of these trials with a logical error.

Logical errors were identified using \textit{encoded Bell pairs}. In particular, each of the code’s $k$ logical qubits was initially entangled with a reference qubit, giving the state

\begin{equation}
    \frac{1}{\sqrt{2^k}}
      \sum_{x\in\{0,1\}^k}\ket{x}_L\ket{x},
\end{equation}

\noindent where $\ket{x}_L$ denotes an encoded computational-basis state of the given code. Noise and syndrome extraction were applied only to this encoded system; the reference states $\ket{x}$ were kept noiseless. A trial had a logical error if, after both corrections, $\bar X_j\otimes X_j$ or $\bar Z_j\otimes Z_j$ had eigenvalue $-1$ for any $j\in\{1,\dots,k\}$.

For Shor’s procedure and ours, complete syndrome measurements continued until $t+1$ consecutive results agreed, or a maximum of $(t+1)^2$ syndrome measurements was reached, where $t$ is the number of errors the code is guaranteed to correct. In the latter case, the last syndrome was used and the shot remained included in the statistics. Steane's procedure instead used only a single complete syndrome measurement. 

This concludes the simulation methodology. As a final note, Figures \ref{fig:unbiased} and \ref{fig:suppdata} show that our procedures generally performed more like Shor syndrome extraction than Steane syndrome extraction, with the separation increasing with $t$. This trend follows from our procedure and Shor's adopting this same $t+1$ repetition protocol, and is not necessarily an intrinsic property of our construction; if simultaneous measurements allow our procedure to use fewer repetitions while preserving fault tolerance, our scheme's performance could move closer to Steane’s.

\begin{figure}
    \centering
    \includegraphics[width=\linewidth]{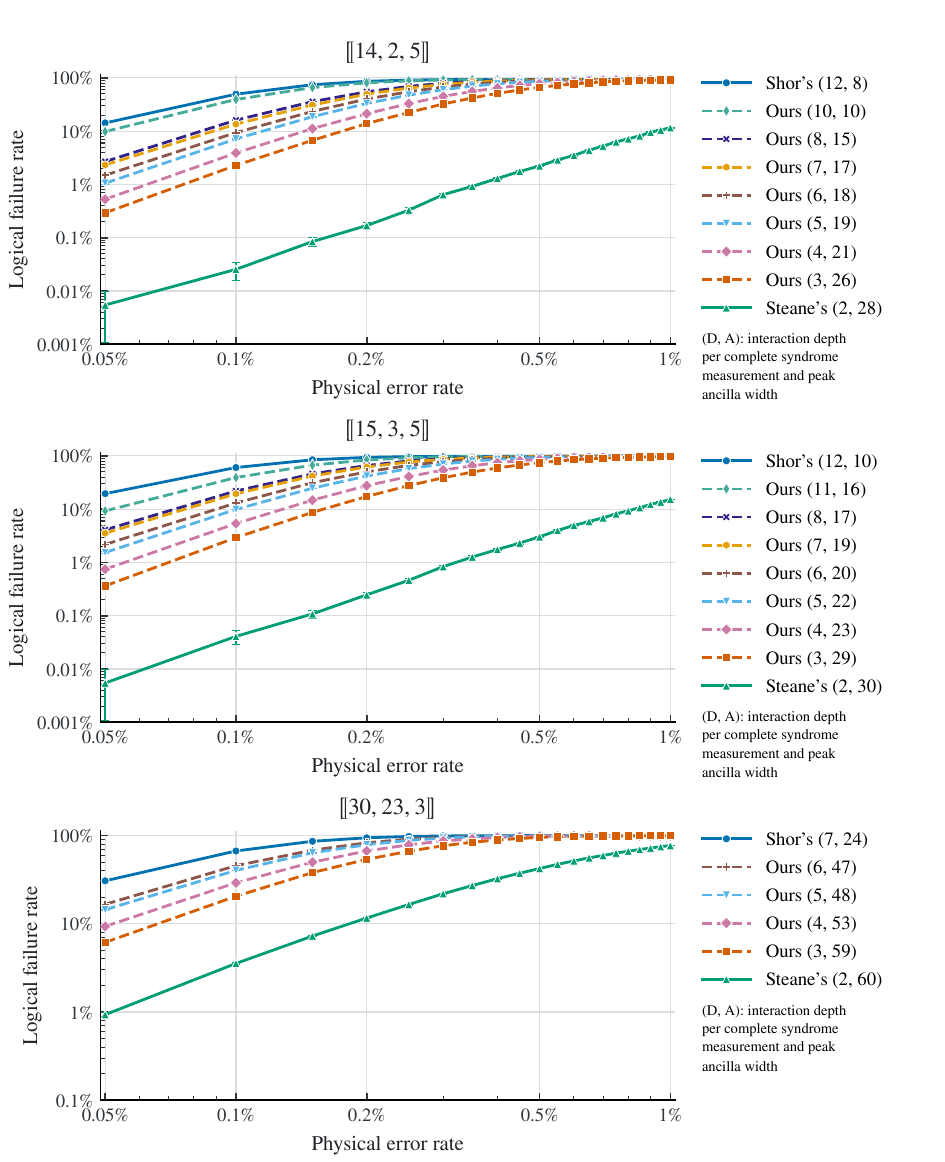}
    \caption{
    Performance of Shor, Steane, and our syndrome extraction circuits for a $\llbracket14,2,5\rrbracket$, $\llbracket15,3,5\rrbracket$, and $\llbracket30,23,3\rrbracket$ code. Plotting conventions are described in the caption of Figure \ref{fig:unbiased}.}
    \label{fig:suppdata}
\end{figure}

\end{appendices}

\end{document}